\documentclass[aps,twocolumn,epsfig,prb]{revtex4}

\newcommand{\bsigma}{\mbox{\boldmath $\sigma$}}

\usepackage{amsmath}
\usepackage{graphicx}

\def\nn{\nonumber}

\begin{document}

\title{Chiral Gauge Theory for Graphene Edge}

\author{Ken-ichi Sasaki}
\email[Email address: ]{SASAKI.Kenichi@nims.go.jp}
\affiliation{International Center for Materials Nanoarchitectonics, 
National Institute for Materials Science,
Namiki, Tsukuba 305-0044, Japan}

\author{Katsunori Wakabayashi}
\affiliation{International Center for Materials Nanoarchitectonics, 
National Institute for Materials Science,
Namiki, Tsukuba 305-0044, Japan}
\affiliation{PRESTO, Japan Science and Technology Agency,
Kawaguchi 332-0012, Japan}

\date{\today}
 
\begin{abstract}
 An effective-mass theory with a deformation-induced (an axial) gauge
 field is proposed as a theoretical framework to study graphene edge.
 Though the gauge field is singular at edge, 
 it can represent the boundary condition and 
 this framework is adopted to solve the scattering problems
 for the zigzag and armchair edges.
 Furthermore, we solve the scattering problem 
 in the presence of a mass term and an electromagnetic field.
 It is shown that 
 the mass term makes the standing wave at the Dirac point 
 avoid the zigzag edge, by which the local density of states
 disappears, 
 and the lowest and first Landau states are special near the zigzag edge.
 The (chiral) gauge theory framework 
 provides a useful description of graphene edge.
\end{abstract}

\maketitle

\section{Introduction}

The graphene edge has attracted much
attention,~\cite{kosynkin09,jiao09,jia09,girit09,liu09,stampfer09,han10,gallagher10} 
because it is the source of a wide variety of notable phenomena. 
For example, the zigzag edge possesses localized edge states.~\cite{tanaka87,kobayashi93,fujita96,nakada96}
The edge states enhance the local density of states
near the Fermi energy.~\cite{klusek00,giunta01,kobayashi05,niimi05} 
As a result, the spins of the edge states may be polarized
by coulombic interaction.~\cite{fujita96}
Another type of edge, the armchair edge, 
does not support edge states.
The zigzag edge is the source of intravalley scattering, 
while the armchair edge gives rise to intervalley scattering.
The transport properties near the armchair edge 
may differ significantly from that near the zigzag edge;~\cite{wakabayashi07,yamamoto09}
however, the reason for this variety is unclear.

The Schr\"odinger equation is a differential equation; therefore, 
an appropriate boundary condition should be imposed on the equation.
The boundary condition is sensitive to the situation of the edge, while 
the local dynamics, as described by the Schr\"odinger equation, 
are the same everywhere in a graphene sample.
The wave function and energy spectrum 
are dependent on the boundary condition.
In this sense, 
the boundary condition is the origin of the variety.~\cite{berry87,mccann04,akhmerov07,akhmerov08}
In this paper,
we attempt to construct a theoretical framework
in which the edge is taken into account as a gauge field,
and not as a boundary condition for the wave function. 
We show that the framework is useful 
to obtain and understand 
the standing wave and edge states.

This paper is organized as follows.
In Sec.~\ref{sec:pspin}, 
the qualitative features of the reflections 
from the zigzag and armchair edges are shown using the kinematics for elastic scattering.
In Sec.~\ref{sec:model}
a general form of the electronic Hamiltonian 
is given for a graphene sheet with edges.
In Secs.~\ref{sec:zig} and \ref{sec:arm},
the scattering problem is solved for both the zigzag and armchair edges, and the standing wave solution is obtained.
A discussion and summary are given 
in Sec.~\ref{sec:dis}.

\section{Reflection of Pseudospin}\label{sec:pspin}

\begin{figure}[htbp]
 \begin{center}
  \includegraphics[scale=0.6]{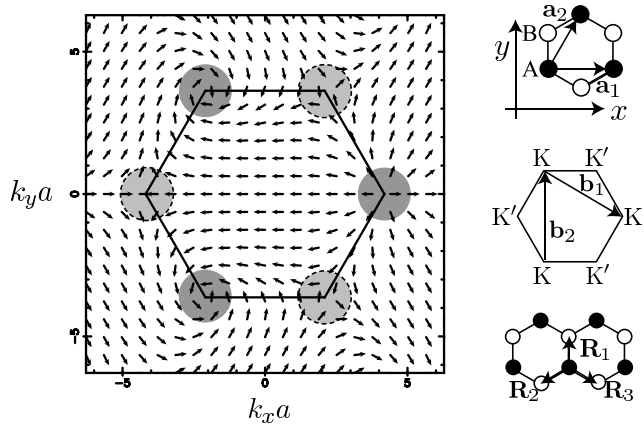}
 \end{center}
 \caption{
 The pseudospin vector field in graphene BZ. 
 Note that this field is for the conduction band and 
 the pseudospin field for the valence band is given by reversing the
 direction of each arrow.
 The singularities in this pseudospin field 
 correspond to the K or K$'$ points. 
 Appendix~\ref{app:rot-pspin} 
 outlines why the pseudospins at the (three)
 equivalent K (K$'$) points are not identical. 
 [inset-top]
 The hexagonal unit cell of graphene consists of 
 {\rm A} (solid circle) and {\rm B} (open circle) atoms. 
 The $xy$ coordinate system is fixed as shown.
 The vectors ${\bf a}_1$ and ${\bf a}_2$ are primitive translations.
 The length of each of these is $a$ ($a\equiv\sqrt{3}a_{\rm cc}$,
 where $a_{\rm cc}$ is the C-C bond length).
 [inset-middle]
 The vectors ${\bf b}_1$ and ${\bf b}_2$ are reciprocal lattice vectors
 defined by ${\bf a}_i \cdot {\bf b}_j = 2\pi \delta_{ij}$.
 [inset-bottom]
 The vectors ${\bf R}_a$ are expressed as
 ${\bf R}_1=a_{\rm cc}{\bf e}_y$,
 ${\bf R}_2=-(\sqrt{3}/2)a_{\rm cc}{\bf e}_x -(1/2)a_{\rm cc}{\bf e}_y$,
 and ${\bf R}_3=(\sqrt{3}/2)a_{\rm cc}{\bf e}_x -(1/2)a_{\rm cc}{\bf e}_y$,
 where ${\bf e}_x$ (${\bf e}_y$)
 is the dimensionless unit vector for the $x$-axis ($y$-axis).
 }
 \label{fig:pspin}
\end{figure}

In the inset of Fig.~\ref{fig:pspin},
we consider the zigzag edge parallel to the $x$-axis, 
by which translational symmetry along the $y$-axis is broken.
Thus, the incident state with wave vector $(k_x,k_y)$ 
is elastically scattered by the zigzag edge, 
and the wave vector of the reflected state becomes $(k_x,-k_y)$.
In contrast,
the armchair edge parallel to the $y$-axis 
breaks the translational symmetry along the $x$-axis, 
so that the wave vector of the reflected state is $(-k_x,k_y)$.
The Brillouin zone (BZ) 
is given by 90$^\circ$ rotation of the hexagonal lattice,
so that for the incident state near the K point in Fig.~\ref{fig:pspin}, 
the zigzag edge reflected state is also near the K point, while
the armchair edge reflected state is near the K$'$ point.  
Therefore, scattering by the zigzag edge is intravalley scattering, 
while that by the armchair edge is intervalley scattering. 


Pseudospin is defined as 
the expected value of the Pauli matrices $\sigma_{x,y,z}$
with respect to the two component Bloch function.
The pseudospin provides information concerning 
the relative phase and the relative amplitude
between the two components of the Bloch function,
and it can be used to 
characterize scattering at the edges.~\cite{sasaki10-jpsj}
The Bloch function of the conduction state 
with wave vector ${\bf k}=(k_x,k_y)$ is given by 
\begin{align}
 \Psi^c_{\bf k} = \frac{1}{\sqrt{2}}
 \begin{pmatrix}
  1 \cr - \frac{f_{\bf k}^*}{|f_{\bf k}|}
 \end{pmatrix},
 \label{eq:bloch}
\end{align}
where $f_{\bf k} = \sum_a e^{i{\bf k}\cdot {\bf R}_a}$,
$f_{\bf k}^*$ denotes the complex conjugate of $f_{\bf k}$,
and ${\bf R}_a$ ($a=1,2,3$) are the vectors 
pointing to the nearest-neighbor {\rm B} atoms 
from an {\rm A} atom [see the inset of Fig.~\ref{fig:pspin}].
The pseudospin is then given by
\begin{align}
 \langle \sigma_x \rangle_{\bf k} = - \frac{{\rm Re}[f_{\bf k}]}{|f_{\bf k}|}, \ \ 
 \langle \sigma_y \rangle_{\bf k} = \frac{{\rm Im}[f_{\bf k}]}{|f_{\bf k}|}, \ \
 \langle \sigma_z \rangle_{\bf k} = 0,
 \label{eq:pspin2D}
\end{align}
where 
\begin{align}
\begin{split}
 & {\rm Re}[f_{\bf k}] = \cos \left(\frac{k_y a}{\sqrt{3}} \right) + 
 2\cos\left(\frac{k_y a}{2\sqrt{3}}\right) \cos\left(\frac{k_x
 a}{2}\right),
 \\
 & {\rm Im}[f_{\bf k}] = \sin \left(\frac{k_y a}{\sqrt{3}} \right) -
 2\sin\left(\frac{k_y a}{2\sqrt{3}}\right) \cos\left(\frac{k_x
 a}{2}\right).
\end{split} 
 \label{eq:ref-imf}
\end{align}
The pseudospin,
$(\langle \sigma_x \rangle_{\bf k},\langle \sigma_y \rangle_{\bf k},\langle \sigma_z \rangle_{\bf k})$,
may be regarded as a two-dimensional vector field,
because $\langle \sigma_z \rangle_{\bf k} =0$.
The arrows in Fig.~\ref{fig:pspin} show the pseudospin field,
$(\langle \sigma_x \rangle_{\bf k},\langle \sigma_y \rangle_{\bf k})$.
$\langle \sigma_y \rangle_{\bf k}$ is proportional to ${\rm Im}[f_{\bf k}]$; therefore,
the angle of each arrow with respect to the $k_x$-axis
represents the relative phase of the Bloch function between A and B atoms.
For example, at the $\Gamma$ point ${\bf k}=0$ in Fig.~\ref{fig:pspin}, 
the arrow is pointing toward the negative $k_x$-axis.
This implies that the wave function forms 
an antisymmetric combination 
with respect to the A and B atoms,
which can be checked by setting ${\bf k}=0$ in Eq.~(\ref{eq:bloch}).
Since $\langle \sigma_y \rangle_{\bf k}$ is an odd function of
$k_y$ as shown in Eq.~(\ref{eq:ref-imf}), 
the pseudospin in Fig.~\ref{fig:pspin} at $(k_x,k_y)$ 
and that at $(k_x,-k_y)$ 
point to different orientations with respect to $\langle \sigma_y \rangle$.
In contrast, the pseudospin at $(k_x,k_y)$ and 
that at $(-k_x,k_y)$ point toward the same orientation.
Thus, the pseudospin component perpendicular to the zigzag edge flips, 
while the pseudospin is invariant for the armchair edge.

We have seen for the zigzag edge that 
the reflection is intravalley scattering and that 
the pseudospin component perpendicular to the edge flips.
For the armchair edge, 
the reflection is intervalley scattering
and the pseudospin is invariant. 
More details concerning the scattering, for example, 
the relative phase between the incident and reflected waves
and the edge states are difficult to obtain 
within the above argument. 
In subsequent sections 
we will explore an effective Hamiltonian
to obtain the standing wave and the edge states.

\section{Deformation-induced gauge field}\label{sec:model}

The fact that the pseudospin flips at the zigzag edge
leads us to consider a gauge field for the edge 
that couples with the pseudospin in a manner similar to that 
an electromagnetic gauge field couples with the real spin.
Here, we show the formulation, in which the effect of the edge
is included into the Hamiltonian as a deformation-induced gauge field.~\cite{sasaki08ptps}

To begin with, we consider 
a change of the nearest-neighbor hopping integral 
from the average value, $-\gamma_0$, as 
$-\gamma_0 + \delta \gamma_{0,a}({\bf r})$,
where $a$ $(=1,2,3)$ denotes the direction of a bond
parallel to ${\bf R}_a$ in the inset of Fig.~\ref{fig:pspin}.
The deviation $\delta \gamma_{0,a}({\bf r})$
represents a lattice deformation in a graphene sheet.
The low energy effective-mass equation 
for deformed graphene is written as 
\begin{align}
 H({\bf r})
 \begin{pmatrix}
  \Psi_{\rm K}({\bf r}) \cr \Psi_{\rm K'}({\bf r})
 \end{pmatrix}
 = E
 \begin{pmatrix}
  \Psi_{\rm K}({\bf r}) \cr \Psi_{\rm K'}({\bf r})
 \end{pmatrix},
 \label{eq:hwave}
\end{align}
where 
$\Psi_{\rm K}({\bf r})$ and $\Psi_{\rm K'}({\bf r})$ 
are two-component wavefunctions 
that represent the electrons near the K and K$'$ points, 
respectively. 
The Hamiltonian for deformed graphene is written as~\cite{sasaki08ptps}
\begin{align}
 H({\bf r})
 = v_{\rm F}
 \begin{pmatrix}
  \bsigma \cdot ({\bf {\hat p}}+{\bf A}^{\rm q}({\bf r})) & 
  \phi^{\rm q}({\bf r}) \sigma_x \cr
  \phi^{\rm q}({\bf r})^* \sigma_x & 
  \bsigma' \cdot ({\bf {\hat p}}-{\bf A}^{\rm q}({\bf r}))
 \end{pmatrix},
 \label{eq:H}
\end{align}
where ${\hat {\bf p}}=-i\hbar \nabla$
is the momentum operator,
$\bsigma=(\sigma_x,\sigma_y)$, and
$\bsigma'=(-\sigma_x,\sigma_y)$.
A lattice deformation $\delta \gamma_{0,a}({\bf r})$
enters the Hamiltonian through the deformation-induced gauge field
${\bf A}^{\rm q}({\bf r})=(A_x^{\rm q}({\bf r}), A_y^{\rm q}({\bf r}))$,
where
${\bf A}^{\rm q}({\bf r})$ is expressed by a linear combination of 
$\delta \gamma_{0,a}(\bf{r})$ as~\cite{kane97,sasaki08ptps,katsnelson08}
\begin{align}
 \begin{split}
  & v_{\rm F} A^{\rm q}_x({\bf r}) = \delta \gamma_{0,1}({\bf r}) 
  - \frac{1}{2} \left(
  \delta \gamma_{0,2}({\bf r})+ \delta \gamma_{0,3}({\bf r}) \right), \\
  & v_{\rm F} A^{\rm q}_y({\bf r}) = \frac{\sqrt{3}}{2} 
  \left( \delta \gamma_{0,2}({\bf r})- \delta \gamma_{0,3}({\bf r}) \right).
 \end{split}
 \label{eq:gauge}
\end{align}
The ${\bf A}^{\rm q}({\bf r})$ field 
causes intravalley scattering, while 
the perturbation that is relevant to intervalley scattering 
is given by a linear combination of $A_x^{\rm q}({\bf r})$
and $A_y^{\rm q}({\bf r})$ as~\cite{sasaki08ptps}
\begin{align}
 \phi^{\rm q}({\bf r}) 
 \equiv (A_x^{\rm q}({\bf r})+iA_y^{\rm q}({\bf r}))
 e^{-2ik_{\rm F}x}.
 \label{eq:gauge2}
\end{align}

\begin{figure}[htbp]
 \begin{center}
  \includegraphics[scale=0.5]{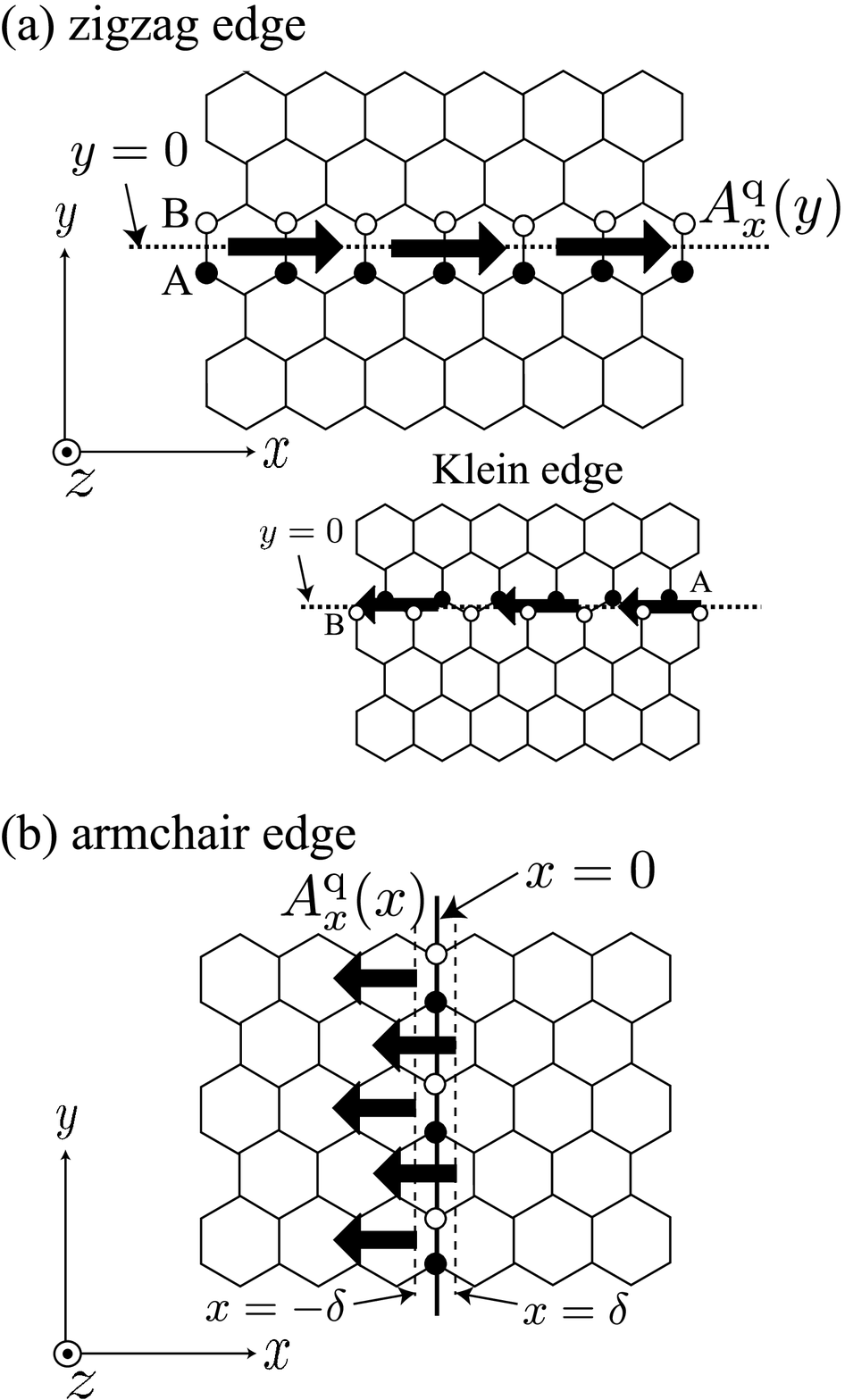}
 \end{center}
 \caption{(a) 
 The bonds on the dotted line at $y=0$ are cut to introduce the
 zigzag edge (Klein edge). 
 The cutting is represented as a deformation-induced gauge field 
 ${\bf A}^{\rm q}({\bf r}) =(A_x^{\rm q}(y),0)$.
 (b) The deformation-induced gauge field for the armchair edge is given by
 ${\bf A}^{\rm q}({\bf r}) =(A_x^{\rm q}(x),0)$.
 }
 \label{fig:graphene}
\end{figure}

In Fig.~\ref{fig:graphene}(a),
we consider cutting the C-C bonds located on the $x$-axis at $y=0$
in order to introduce the zigzag edge 
in a flat graphene sheet.
After cutting the bonds, 
the graphene sheet splits into two semi-infinite parts:
$y>0$ and $y<0$.
The cutting is represented as 
$\delta \gamma_{0,1}({\bf r})|_{y=0}=\gamma_0$,
$\delta \gamma_{0,2}({\bf r})=0$ and
$\delta \gamma_{0,3}({\bf r})=0$.
From Eq.~(\ref{eq:gauge}), 
the corresponding deformation-induced gauge field
is then written as
${\bf A}^{\rm q}({\bf r}) =(A_x^{\rm q}(y),0)$,
where $A_x^{\rm q}(y)$ is not vanishing 
only for the C-C bonds located on the $x$-axis at $y=0$ 
as $A_x^{\rm q}(y)|_{y=0}=(\gamma_0/v_{\rm F})$.
Since $A_x^{\rm q}(y)$ is defined for the C-C bond,
$A_x^{\rm q}(y)$ is meaningful when it is integrated 
from $-\xi_g$ to $\xi_g$, where 
$\xi_g$ is of the same order as the C-C bond length
and will be taken to be zero at the end of calculation in the continuum
limit. 
Note also that 
the vector direction of ${\bf A}^{\rm q}({\bf r})$ 
is perpendicular to that of the bond 
with a modified hopping integral.
Since the zigzag edge is not the source of intervalley scattering,
intervalley scattering can be ignored. 
Hereafter, we consider the electrons near the K point for the zigzag edge.
Moreover, separation of variables can be employed, 
due to translational symmetry along the $x$-axis.
As a result, 
$\Psi_{\rm K}({\bf r})$ and $\hat{p}_x$ in Eq.~(\ref{eq:H})
can be replaced with $e^{ik_x x}\Psi_{\rm K}(y)$ and $p_x$.
The energy eigenequation can then be simplified as
$H_{\rm K}(y) \Psi_{\rm K}(y) = E \Psi_{\rm K}(y)$,
where the Hamiltonian is
\begin{align}
 H_{\rm K}(y) \equiv v_{\rm F} \left[
 \sigma_x (p_x + A_x^{\rm q}(y)) + \sigma_y \hat{p}_y
 \right].
 \label{eq:HK_zigzag}
\end{align}
This Hamiltonian is solved in Sec.~\ref{sec:zig}.
The cutting which produces the Klein edges~\cite{klein94,klein99} 
is represented as 
$\delta \gamma_{0,1}({\bf r})=0$,
$\delta \gamma_{0,2}({\bf r})|_{y=0}=\gamma_0$ and
$\delta \gamma_{0,3}({\bf r})|_{y=0}=\gamma_0$.
From Eq.~(\ref{eq:gauge}), 
the corresponding deformation-induced gauge field
is then written as
${\bf A}^{\rm q}({\bf r}) =(-A_x^{\rm q}(y),0)$,
where $A_x^{\rm q}(y)$ is the gauge field for the zigzag edge. 
Note that the direction of the ${\bf A}^{\rm q}({\bf r})$ field 
for the Klein edge is opposite that of the zigzag edge.

The armchair edge can be introduced by 
cutting the bonds located on $x=\pm \delta$,
as shown in Fig.~\ref{fig:graphene}(b).
By setting 
$\delta \gamma_{0,1}({\bf r})=0$,
$\delta \gamma_{0,2}({\bf r})|_{x=-\delta}=\gamma_0$ and
$\delta \gamma_{0,3}({\bf r})|_{x=\delta}=\gamma_0$
in Eq.~(\ref{eq:gauge}), 
the deformation-induced gauge field 
for the armchair edge is written as 
${\bf A}^{\rm q}({\bf r})=(A_x^{\rm q}(x),0)$
with the limit of $\delta \to 0$.
Due to translational symmetry along the $y$-axis, 
$\hat{p}_y$ is replaced to $p_y$ in Eq.~(\ref{eq:H}).
Thus, the Hamiltonian is given by 
\begin{align}
 v_{\rm F}
 \begin{pmatrix}
  \sigma_x (\hat{p}_x+A_x^{\rm q}(x)) + \sigma_y p_y & 
  \phi^{\rm q}(x) \sigma_x \cr
  \phi^{\rm q}(x)^* \sigma_x & 
  -\sigma_x (\hat{p}_x-A_x^{\rm q}(x)) + \sigma_y p_y
 \end{pmatrix}, \nn
\end{align}
with $\phi^{\rm q}(x) = A^{\rm q}_x(x) e^{-2ik_{\rm F} x}$.
This Hamiltonian can be reduced further by means of
a gauge symmetry in the following manner.
Since $A_x^{\rm q}(x)$ does not depend on $y$, 
it can be represented in terms of a scalar function $\varphi(x)$, as
$A_x^{\rm q}(x) = \partial_x \varphi(x)$.
Using the gauge transformation: 
$\Psi_{\rm K}(x) \to e^{-i\varphi(x)} \Psi_{\rm K}(x)$
and $\Psi_{\rm K'}(x) \to e^{i\varphi(x)} \Psi_{\rm K'}(x)$,
$A_x^{\rm q}(x)$ can be erased from the Hamiltonian for each valley.
However, note that as a result of this gauge transformation, 
$\phi^{\rm q}(x)$ must be changed into $e^{2i\varphi(x)} \phi^{\rm q}(x)$.
To minimize notation,
let us use $\phi^{\rm q}(x)$ to denote this gauge transformed field,
so that 
$\phi^{\rm q}(x)\equiv A^{\rm q}_x(x) e^{2i[\varphi(x)-k_{\rm F}x]}$.
The Hamiltonian for the armchair edge is then written as
\begin{align}
 H(x) = v_{\rm F}
 \begin{pmatrix}
  \sigma_x \hat{p}_x + \sigma_y p_y & 
  \phi^{\rm q}(x) \sigma_x \cr
  \phi^{\rm q}(x)^* \sigma_x & 
  -\sigma_x \hat{p}_x + \sigma_y p_y
 \end{pmatrix}. 
 \label{eq:armchairH}
\end{align}
This will be solved in Sec.~\ref{sec:arm}.
Note that by introducing $\tau_\alpha$ ($\alpha=1,2,3$) matrices 
defined by
\begin{align}
 \tau_1 = 
  \begin{pmatrix}
  0 & I \cr
  I & 0 
 \end{pmatrix}, \
 \tau_2 = 
 \begin{pmatrix}
  0 & -iI \cr
  iI & 0 
 \end{pmatrix}, \
  \tau_3 = 
  \begin{pmatrix}
  I & 0 \cr
  0 & -I 
 \end{pmatrix},
\end{align}
the unperturbed Hamiltonian is represented in a compact fashion as
$H_0({\bf r})=v_{\rm F}(\tau_3 \sigma_x \hat{p}_x + \tau_0 \sigma_y \hat{p}_y)$,
where $\tau_0$ is a $4\times 4$ identity matrix.

In Eqs.~(\ref{eq:gauge}) and (\ref{eq:gauge2}),
we assume $|\delta \gamma_{0,a}({\bf r})| \ll \gamma_0$,
and ignore the higher order term of $\delta \gamma_{0,a}({\bf r})$.
As a result of this simplification, 
the relationship between ${\bf A}^{\rm q}({\bf r})$ 
and $\delta \gamma_{0,a}({\bf r})$ may deviate from Eq.~(\ref{eq:gauge})
when $|\delta \gamma_{0,a}(\mathbf{r})|\approx \gamma_0$.
However, note that the direction and not the strength 
of the ${\bf A}^{\rm q}({\bf r})$ field can be determined by
Eq.~(\ref{eq:gauge}), 
even for the case where $|\delta \gamma_{0,a}({\bf r})|\approx \gamma_0$.
Consideration of this point is given in Appendix~\ref{app:rot-pspin}.

\section{Zigzag edge}\label{sec:zig}

The scattering problem for the zigzag edge is solved in this section.
Standing wave solutions are constructed in Sec.~\ref{sec:exs},
and the properties of the solutions are examined in detail. 
Localized edge states are constructed in Sec.~\ref{ssec:es}.
The behavior of the standing wave in the presence of a mass term 
and an external magnetic field is examined in Secs.~\ref{ssec:mass}
and~\ref{ssec:emf}, respectively.
The local density of states near the zigzag edge is calculated
analytically in Sec.~\ref{ssec:ldos}.

\subsection{Standing Waves}\label{sec:exs}

To begin with, 
solutions are constructed for the case of $A_x^{\rm q}(y)=0$
in Eq.~(\ref{eq:HK_zigzag}).
Let $\Phi(y)$ be the eigenstate of the unperturbed Hamiltonian
$H_{\rm K}^0(y) = v_{\rm F} \left( \sigma_x p_x + \sigma_y \hat{p}_y \right)$.
$H_{\rm K}^0(y)$ satisfies
$\sigma_x H_{\rm K}^0(-y) \sigma_x = H_{\rm K}^0(y)$; therefore, 
a general solution may be constructed from the basis function 
$\Phi(y)$ to satisfy the constraint equation,
\begin{align}
 \Phi(-y) = e^{-ig} \sigma_x \Phi(y),
 \label{eq:sym}
\end{align}
where $g$ is a real number phase.
The phase $g$ can not be an arbitrary value.
The successive operation of Eq.~(\ref{eq:sym}) on $\Phi(-y)$
gives $\Phi(-(-y))=e^{-2ig} \sigma_x^2 \Phi(y)$,
and hence $g$ should be $0$ or $\pi$.
Note that a set of functions satisfying Eq.~(\ref{eq:sym})
is useful for construction of solutions in the case $A_x^{\rm q}(y)\ne 0$,
because $H_{\rm K}(y)$ also satisfies 
$\sigma_x H_{\rm K}(-y) \sigma_x = H_{\rm K}(y)$.
This constraint comes from the inversion symmetry 
of the gauge field with respect to $y=0$,
$A_x^{\rm q}(-y) = A_x^{\rm q}(y)$.

From Eq.~(\ref{eq:sym}), we have
$\Phi_{\rm B}(y)=e^{ig} \Phi_{\rm A}(-y)$.
Thus, $\Phi(y)$ can be rewritten as
\begin{align}
 \Phi(y) = 
 \begin{pmatrix}
  \Phi_{\rm A}(y) \cr \Phi_{\rm B}(y)
 \end{pmatrix}
 =
 \begin{pmatrix}
  \Phi_{\rm A}(y) \cr e^{ig} \Phi_{\rm A}(-y)
 \end{pmatrix}.
 \label{eq:wf-zig}
\end{align}
By substituting Eq.~(\ref{eq:wf-zig}) into
$H_{\rm K}^0(y) \Phi(y)=E \Phi(y)$,
we obtain simultaneous differential equations:
\begin{align}
\begin{split}
 & \frac{E}{v_{\rm F}} \Phi_s(y)
 = + p_x \Phi_s(y)
 + \hbar \frac{d}{dy} \Phi_a(y), \\
 & \frac{E}{v_{\rm F}} \Phi_a(y)
 = - p_x \Phi_a(y)
 - \hbar \frac{d}{dy} \Phi_s(y),
\end{split}
 \label{eq:Hsym}
\end{align}
where $\Phi_s(y)$ and $\Phi_a(y)$ are defined as
\begin{align}
\begin{split}
 & \Phi_s(y) \equiv 
 e^{-i\frac{g}{2}} \Phi_{\rm A}(y) + 
 e^{+i\frac{g}{2}} \Phi_{\rm A}(-y), \\
 & \Phi_a(y) \equiv 
 e^{-i\frac{g}{2}} \Phi_{\rm A}(y) -
 e^{+i\frac{g}{2}} \Phi_{\rm A}(-y).
\end{split}
\label{eq:PhiAB-def}
\end{align}
For the case $g=0$, 
Eq.~(\ref{eq:PhiAB-def}) implies that 
$\Phi_s(y)$ is an even function [$\Phi_s(y)=\Phi_s(-y)$], 
while $\Phi_a(y)$ is an odd function [$\Phi_a(y)=-\Phi_a(-y)$].
Thus, they can be parameterized as follows:
\begin{align}
\begin{split}
 & \Phi_s(y) = S \cos (k_y y), \\
 & \Phi_a(y) = A \sin (k_y y),
\end{split}
\label{eq:func0}
\end{align}
where the parameters $S$ and $A$ 
can be determined from Eq.~(\ref{eq:Hsym}).
By substituting Eq.~({\ref{eq:func0}}) into Eq.~(\ref{eq:Hsym}),
we obtain the secular equation
\begin{align}
 \begin{pmatrix}
  \frac{E}{\hbar v_{\rm F}}-k_x & -k_y \cr
  -k_y & \frac{E}{\hbar v_{\rm F}}+k_x
 \end{pmatrix}
\begin{pmatrix}
 S \cr A
\end{pmatrix}
 = 0.
\end{align}
The solution of this secular equation satisfies
\begin{align}
\begin{split}
 & E^2 = (\hbar v_F)^2 (k_x^2 + k_y^2), \\
 & A = \frac{k_y}{\frac{E}{\hbar v_{\rm F}}+k_x} S.
\end{split}
\label{eq:eneA}
\end{align}
Let $\theta({\bf k})$ be the polar angle 
between vector ${\bf k}$ and the $k_x$-axis. 
Then, $k_x = k\cos \theta({\bf k})$ and $k_y = k\sin \theta({\bf k})$ 
where $k=|{\bf k}|$, and 
the second equation of Eq.~(\ref{eq:eneA})
can be rewritten as 
$A = S \tan \left[ \theta({\bf k})/2 \right]$ 
for the eigenstate with positive energy $E=\hbar v_{\rm F}k$.
Assuming that $S=\cos \left[ \theta({\bf k})/2 \right]$, 
we have $A=\sin \left[ \theta({\bf k})/2 \right]$.
Substituting these into Eq.~(\ref{eq:func0}) gives
\begin{align}
\begin{split}
 & \Phi_s(y) = \cos \left( \frac{\theta({\bf k})}{2} \right) \cos (k_y y), \\
 & \Phi_a(y) = \sin \left( \frac{\theta({\bf k})}{2} \right) \sin (k_y y).
\end{split}
\label{eq:phisa}
\end{align}
Then, Eq.~(\ref{eq:phisa}) is substituted into Eq.~(\ref{eq:PhiAB-def}) with $g=0$ to give 
\begin{align}
 \Phi^{0}(y) =
\begin{pmatrix}
 \cos \left( k_y y - \frac{\theta({\bf k})}{2} \right) \cr 
 \cos \left( k_y y + \frac{\theta({\bf k})}{2} \right)
\end{pmatrix}.
\label{eq:phig0}
\end{align}
Similarly, for the case where $g=\pi$, we have
\begin{align}
 \Phi^{\pi}(y) = 
\begin{pmatrix}
 \sin \left( k_y y - \frac{\theta({\bf k})}{2} \right) \cr
 \sin \left( k_y y + \frac{\theta({\bf k})}{2} \right)
\end{pmatrix}.
\label{eq:phigpi}
\end{align}
The energies of the eigenstates 
$\Phi^{0}(y)$ and $\Phi^{\pi}(y)$ are equal, and therefore
a general solution can be expressed as a 
superposition of the degenerate eigenstates, as
\begin{align}
 \Phi^{f}(y) 
 &\equiv \sin(f) \Phi^{0}(y) 
 + \cos(f) \Phi^{\pi}(y) \nonumber \\
 &= \begin{pmatrix}
  \sin \left( k_y y - \theta({\bf k})/2 + f \right) \cr
  \sin \left( k_y y + \theta({\bf k})/2 + f \right)
 \end{pmatrix},
 \label{eq:phify}
\end{align}
where $f$ is a real number.
The value of $f$ is determined as follows.

The Hamiltonian, $H_{\rm K}(y)=H_{\rm K}^0(y)+v_{\rm F}\sigma_x A^{\rm q}_x(y)$, 
is identical to the unperturbed Hamiltonian $H_{\rm K}^0(y)$ 
for $y$ to satisfy $|y| \ge \xi_g$, so that 
$\Phi^{f}(y)$ satisfies the eigenequation
$H_{\rm K}(y) \Phi^f(y) = E \Phi^f(y)$ for $|y| \ge \xi_g$.
We need to solve 
$H_{\rm K}(y) \Psi_{\rm K}(y) = E \Psi_{\rm K}(y)$ 
locally for $|y| < \xi_g$.
By parameterizing the eigenstate of $H_{\rm K}(y)$
as $\Psi_{\rm K}(y)=N(y)\Phi^f(y)$,
we obtain the constraint equation for $N(y)$ and $\Phi^{f}(y)$ as
\begin{align}
 \left\{ \sigma_y [\hat{p}_y N(y)] + \sigma_x A_x^{\rm q}(y)N(y) \right\}
 \Phi^{f}(y) =0.
 \label{eq:constz}
\end{align}
To obtain Eq.~(\ref{eq:constz})
we must place $\Psi_{\rm K}(y)=N(y)\Phi^f(y)$ to 
$H_{\rm K}(y) \Psi_{\rm K}(y) = E \Psi_{\rm K}(y)$,
and use $H_{\rm K}^0(y) \Phi^{f}(y) = E \Phi^{f}(y)$.
Here, we have assumed that 
the energy eigenvalues of the standing wave 
$\Psi_{\rm K}(y)$ and of $\Phi^{f}(y)$
are the same.
This assumption is valid for the standing wave,
because the energy eigenvalue is determined by the bulk Hamiltonian 
$H_{\rm K}^0(y)$ and the energy does not change through 
elastic scattering.
However, note that this assumption is not valid
for the edge states, which  
$H_{\rm K}(y) \Psi_{\rm K}(y) = E \Psi_{\rm K}(y)$ must be solved directly
(see Sec.~\ref{ssec:es} for more details).
Now, Eq.~(\ref{eq:constz}) is equivalent to 
the two successive equations: 
\begin{align}
\begin{split}
 & \left( A_x^{\rm q}(y)N(y) - \hbar \frac{dN(y)}{dy} \right)
 \Phi^{f}_{\rm B}(y) = 0, \\
 & \left( A_x^{\rm q}(y)N(y) + \hbar \frac{dN(y)}{dy} \right)
 \Phi^{f}_{\rm A}(y) = 0.
\end{split}
\label{eq:caseEq}
\end{align}
The following two cases 
can be considered for this successive equation. 
One case is that the solution satisfies
\begin{align}
\begin{split}
 & A_x^{\rm q}(y)N(y) + \hbar \frac{dN(y)}{dy} = 0, \\
 & \Phi^{f}_{\rm B}(y) = 0,
\end{split} \ \ \
 (|y|\le \xi_g).
 \label{eq:case1}
\end{align}
The first (second) equation of Eq.~(\ref{eq:case1})
ensures the second (first) equation
of Eq.~(\ref{eq:caseEq}).
The other case is that the solution satisfies
\begin{align}
\begin{split}
 & A_x^{\rm q}(y)N(y) - \hbar \frac{dN(y)}{dy} = 0, \\
 & \Phi^{f}_{\rm A}(y) = 0,
\end{split} \ \ \
 (|y|\le \xi_g).
 \label{eq:case2}
\end{align}
The two conditions, Eqs.~(\ref{eq:case1}) and (\ref{eq:case2}),
correspond to the standing wave in the upper semi-infinite graphene
plane for $y>0$ and that in the lower plane for $y<0$
in the limit of $\xi_g=0$, as shown in the following.

For the case of Eq.~(\ref{eq:case1}),
the first equation is integrated with respect to $y$,
to obtain 
\begin{align}
 N(-\xi_g) = N(\xi_g) \exp \left( 
 \frac{1}{\hbar} \int_{-\xi_g}^{\xi_g} A_x^{\rm q}(y)dy
 \right).
 \label{eq:weight}
\end{align}
Hence,
when $(1/\hbar) \int_{-\xi_g}^{\xi_g} A_x^{\rm q}(y)dy \gg 0$, 
$N(\xi_g)$ is negligible compared with $N(-\xi_g)$, and therefore
the standing wave appears only for $y<0$.
In contrast,
when $(1/\hbar) \int_{-\xi_g}^{\xi_g} A_x^{\rm q}(y)dy \ll 0$, 
the standing wave appears only for $y>0$.
The other condition in Eq.~(\ref{eq:case1})
holds for the limit of $\xi_g \to 0$ 
by setting $f=-\theta({\bf k})/2$ in Eq.~(\ref{eq:phify}),
because 
\begin{align}
 \lim_{y\to 0}\Phi^{f=-\theta({\bf k})/2}_{\rm B}(y) = 0. 
 \label{eq:phiB0}
\end{align}
This condition leads to $\Psi_{{\rm K},{\rm B}}(0)=0$, 
which represents the boundary conditions
for the zigzag and Klein edges 
shown in Fig.~\ref{fig:graphene}(a).
Thus, Eq.~(\ref{eq:case1}) covers two situations, depending on the
direction of the gauge field; 
$A_x^{\rm q}(y) \gg 0$ or $A_x^{\rm q}(y) \ll 0$.
That is, when $A_x^{\rm q}(y) \gg 0$, 
Eq.~(\ref{eq:case1}) corresponds to the upper semi-infinite graphene plane 
with the zigzag edge, while when $A_x^{\rm q}(y) \ll 0$, 
Eq.~(\ref{eq:case1}) corresponds to the lower semi-infinite graphene
plane with the Klein edge.
Similarly, when $A_x^{\rm q}(y) \gg 0$, 
Eq.~(\ref{eq:case2}) corresponds to the lower semi-infinite graphene plane 
with the zigzag edge, while when $A_x^{\rm q}(y) \ll 0$, 
Eq.~(\ref{eq:case2}) corresponds to the upper semi-infinite graphene plane 
with the Klein edge.

From Eq.~(\ref{eq:weight}), it follows that 
the gauge field for the edge should be large, 
$\left| (1/\hbar) \int_{-\xi_g}^{\xi_g} A_x^{\rm q}(y)dy \right| \gg 1$.
In Ref.~\onlinecite{sasaki06jpsj}, 
the following was obtained analytically 
\begin{align}
 \frac{1}{\hbar} \int_{-\xi_g}^{\xi_g} A_x^{\rm q}(y)dy = 
 -\ln (1-c),
 \label{eq:gauge-tb}
\end{align}
where $c$ is the parameter that specifies the deformation
as $\delta \gamma_{0,1}({\bf r})|_{y=0}=c \gamma_0$
[see Fig.~\ref{fig:graphene}].
The right-hand side gives logarithmic singularities
for $c=1$ and $c=-\infty$.
The limit $c\to 1$ corresponds to the zigzag edge, while
the limit $c\to -\infty$ represents the Klein edge.
Note that when $c\to -\infty$, the electron is unable to 
have a finite amplitude on the A and B atoms located at $y=0$,
which effectively represents the Klein edge.
Because of the singularity, 
$N(y)$ that satisfies Eq.~(\ref{eq:weight}) 
is similar to the step function;
$N(y)=N \ne 0$ for $y < 0$,
and otherwise $N(y)=0$.

Now, by setting $f=-\theta({\bf k})/2$ in Eq.~(\ref{eq:phify}),
the standing wave in the conduction band is expressed as 
\begin{align}
 \Psi^{c}_{{\rm K},{\bf k}}({\bf r}) = 
 \frac{e^{ik_x x}}{\sqrt{L_x}} N(y)
\begin{pmatrix}
 \sin \left( k_y y - \theta({\bf k}) \right) \cr
 \sin \left( k_y y \right) 
\end{pmatrix},
\label{eq:s-sol}
\end{align}
where the plane wave parallel to the edge with the length $L_x$
is included. 
The standing wave in the valence band 
is obtained by using the particle-hole symmetry of the Hamiltonian,
$\sigma_z H_{\rm K}(y) \sigma_z =-H_{\rm K}(y)$, as
$\Psi^{v}_{{\rm K},{\bf k}}(y) = \sigma_z \Psi^{c}_{{\rm K},{\bf k}}(y)$:
\begin{align}
 \Psi^{v}_{{\rm K},{\bf k}}({\bf r}) =  
 \frac{e^{ik_x x}}{\sqrt{L_x}} N(y)
\begin{pmatrix}
 \sin \left( k_y y - \theta({\bf k}) \right) \cr
 - \sin \left( k_y y \right) 
\end{pmatrix}.
\label{eq:s-solv}
\end{align}

Here, we consider the pseudospin of the standing wave.
The pseudospin for an eigenstate $\Psi(y)$ is defined by
the expected value of the Pauli matrices as 
$\langle \sigma_i \rangle \equiv \int \sigma_i(y) dy$
($i=x,y,z$), where $\sigma_i(y)$ is a pseudospin density
defined by 
$\sigma_i(y) \equiv \Psi^\dagger(y) \sigma_i \Psi(y)$.
Note that the $y$-component of the pseudospin 
is proportional to the imaginary part of the Bloch function, such
as $\sigma_y(y) \propto {\rm Im}[\Psi_{\rm A}^*\Psi_{\rm B}]$.
The Bloch function of the standing wave is real,
so that the $y$-component of the pseudospin for the standing
wave vanishes, that is, $\langle \sigma_y \rangle=0$.
Note also that $\langle \sigma_y \rangle=0$ means that the current
normal to the zigzag edge vanishes. 
It is interesting to note that 
$\langle \sigma_y \rangle=0$ holds 
whenever $\Psi_{{\rm K},{\bf k},{\rm A}}(y)$ and 
$\Psi_{{\rm K},{\bf k},{\rm B}}(y)$ can be taken as real numbers.
This indicates that the result $\langle \sigma_y \rangle=0$
is not sensitive to the value of $f$, but depends only on the fact that
$\Phi^f$ does not have a relative phase between the two components.
The condition of Eq.~(\ref{eq:phiB0}) means that 
the pseudospin density is locally polarized 
into the positive $z$-axis near the zigzag edge, 
that is, $\sigma_z(0) > 0$ and $\sigma_x(0)=\sigma_y(0)=0$.
Actually, by substituting $y \simeq 0$ into Eq.~(\ref{eq:s-sol}),
the standing wave near the zigzag edge
has amplitude only at A-atoms.
This polarization of the pseudospin 
is consistent with the fact that 
the gauge field $A_x^{\rm q}(y)$
has a non-vanishing deformation-induced magnetic field,
\begin{align}
 B_z^{\rm q}({\bf r}) \equiv 
 \partial_x A_y^{\rm q}({\bf r})-\partial_y A_x^{\rm q}({\bf r}),
\end{align}
at the zigzag edge.
The presence of the $B_z^{\rm q}(y)$ field at the zigzag edge 
causes local polarization of the standing wave pseudospin
near the zigzag edge,
similar to the polarization of a real spin by a magnetic field.
We will show in Sec.~\ref{ssec:ldos} that 
this polarization of the pseudospin causes anomalous 
behavior to appear in the local density of states (LDOS)
near the zigzag edge.



A zigzag nanoribbon is given by introducing 
another zigzag edge at $y = -L$, 
in addition to the zigzag edge at $y=0$.
Suppose that the edge atoms at $y = -L$ are B-atoms, 
which imposes the boundary condition on the wave function at $y=-L$
as $\lim_{y=-L}\Psi_{{\rm K},{\bf k},{\rm A}}^c({\bf r}) = 0$.
This leads to the constraint equation for $(k_x,k_y)$,
\begin{align}
 k_y L + \theta({\bf k}) = n\pi,
 \label{eq:const}
\end{align}
where $n$ is an integer.
It is noted that this equation reproduces
\begin{align}
 k_y = -k_x \tan(k_yL), 
 \label{eq:Bery}
\end{align}
which was obtained by Brey and Fertig in Ref.~\onlinecite{brey06}
[the negative sign in front of $k_x$ is a matter of notation].
Note that $n$ should be a nonzero integer, because 
the equation does not possess a solution when $n=0$.
For the case where the edge at $y = -L$ is the Klein edge,
the boundary condition on the wave function at $y=-L$ becomes 
$\lim_{y=-L}\Psi_{{\rm K},{\bf k},{\rm B}}^c({\bf r}) = 0$.
This leads to $k_y L = n\pi$,
where $n$ is a positive integer.

\subsection{Edge States}\label{ssec:es}

In addition to the standing wave derived in the previous subsection, 
$H_{\rm K}(y)$ possesses localized edge states.~\cite{sasaki06jpsj}
Here, we show how to construct the edge states.

The following observation is useful in order to obtain 
the edge states.
Instead of Eq.~(\ref{eq:func0}), 
we assume
\begin{align}
\begin{split}
 & \Phi_s(y) = S \cosh (y/\xi), \\
 & \Phi_a(y) = A \sinh (y/\xi).
\end{split}
 \label{eq:func1}
\end{align}
By substituting Eq.~({\ref{eq:func1}}) into 
Eq.~(\ref{eq:Hsym}), the secular equation is obtained:
\begin{align}
 \begin{pmatrix}
  \frac{E}{\hbar v_{\rm F}}-k_x & -\xi^{-1} \cr
  +\xi^{-1} & \frac{E}{\hbar v_{\rm F}}+k_x
 \end{pmatrix}
\begin{pmatrix}
 S \cr A
\end{pmatrix}
 = 0.
\end{align}
The solution of this secular equation satisfies
\begin{align}
\begin{split}
 & E^2 = (\hbar v_F)^2 \left( k_x^2 - \xi^{-2} \right), \\
 & A = -\frac{\xi^{-1}}{\frac{E}{\hbar v_{\rm F}}+k_x} S.
\end{split}
\end{align}
By introducing the $\phi$ variable, which satisfies
\begin{align}
 \xi^{-1} = - k_x \tanh \phi,
 \label{eq:kycond}
\end{align}
we have $E^2/(\hbar v_F)^2=k_x^2/\cosh^2 \phi$.
For the case  
\begin{align}
 \frac{E}{\hbar v_{\rm F}} = \frac{k_x}{\cosh\phi},
 \label{eq:solu1}
\end{align}
we have $A/S = \tanh (\phi/2)$.
By inserting this into Eq.~(\ref{eq:func1})
and setting $S=\cosh(\phi/2)$,
we obtain 
\begin{align}
\begin{split}
 & \Phi_s(y) = \cosh \left( \frac{\phi}{2} \right) \cosh \left(\frac{y}{\xi}\right), \\
 & \Phi_a(y) = \sinh \left( \frac{\phi}{2} \right) \sinh \left(\frac{y}{\xi}\right).
\end{split}
\label{eq:phisa-e}
\end{align}
By substituting Eq.~(\ref{eq:phisa-e}) into 
Eq.~(\ref{eq:PhiAB-def}) with $g=0$, we have
\begin{align}
 \Phi^{0}(y) = 
\begin{pmatrix}
 \cosh \left( \frac{y}{\xi} + \frac{\phi}{2} \right) \cr 
 \cosh \left( \frac{y}{\xi} - \frac{\phi}{2} \right)
\end{pmatrix}.
\end{align}
Similarly, for the case $g=\pi$, we have
\begin{align}
 \Phi^{\pi}(y) =
\begin{pmatrix}
 \sinh \left( \frac{y}{\xi} + \frac{\phi}{2} \right) \cr
 \sinh \left( \frac{y}{\xi} - \frac{\phi}{2} \right)
\end{pmatrix}.
\end{align}
The energies of $\Phi^{\pi}(y)$ and $\Phi^{0}(y)$
are equal; therefore, 
the basis function may be chosen as
\begin{align}
\begin{split}
 & \Phi^{+}(y) \equiv \Phi^{0}(y) + \Phi^{\pi}(y)
 = e^{+\frac{y}{\xi}} 
 \begin{pmatrix}
  e^{+\phi/2} \cr e^{-\phi/2}
 \end{pmatrix},
 \\
 & \Phi^{-}(y) \equiv \Phi^{0}(y) - \Phi^{\pi}(y)
 = e^{-\frac{y}{\xi}} 
 \begin{pmatrix}
  e^{-\phi/2} \cr e^{+\phi/2}
 \end{pmatrix}.
\end{split} 
 \label{eq:Phi}
\end{align}
The functions $\Phi^+(y)$ and $\Phi^-(y)$ 
are exponentially increasing and decreasing functions of $y$, respectively. 
Thus, neither $\Phi^+(y)$ nor $\Phi^-(y)$ 
is a normalized wave function all over the space, $y\in (-\infty,\infty)$.
However, note that $\Phi^+(y)$ and $\Phi^-(y)$
can be normalizable wave functions for $y<0$ and $y>0$, respectively.
We also note that the pseudospin of $\Phi^+(y)$ 
is given by $\langle \sigma_z \rangle = \tanh\phi$, while 
that of $\Phi^-(y)$ is $\langle \sigma_z \rangle = -\tanh\phi$.

From the above observation, 
we parameterized the localized eigenstate as
\begin{align}
 \Psi_{\rm K}(y) = N e^{-\frac{|y|}{\xi}} 
 \begin{pmatrix}
  e^{+g(y)} \cr e^{-g(y)}
 \end{pmatrix},
 \label{eq:phiR}
\end{align}
where $N$ is a normalization constant, and 
the modulation of the pseudospin is represented by 
a function $g(y)$.
Substituting Eq.~(\ref{eq:phiR}) into 
$H_{\rm K}(y) \Psi_{\rm K}(y) = E \Psi_{\rm K}(y)$ 
gives simultaneous differential equations for $g(y)$,
\begin{align}
 \begin{split}
  & p_x + A^{\rm q}_x(y) +\hbar \frac{d}{dy} \left(\frac{|y|}{\xi}+g(y)\right) =
  \frac{E}{v_F} e^{+2g(y)}, \\ 
  & p_x + A^{\rm q}_x(y) -\hbar \frac{d}{dy} \left(\frac{|y|}{\xi}-g(y)) \right) = 
  \frac{E}{v_F} e^{-2g(y)}.
 \end{split}
 \label{eq:w+g-d}
\end{align}
By summing and subtracting both sides of Eq.~(\ref{eq:w+g-d}), 
the energy eigenequation can be rewritten as 
\begin{align}
 \begin{split}
  & p_x +A^{\rm q}_x(y) + \hbar \frac{dg(y)}{dy} 
  = \frac{E}{v_F} \cosh(2g(y)), \\
  & \hbar \frac{d}{dy}\left(\frac{|y|}{\xi}\right) = \frac{E}{v_F} \sinh(2g(y)).
 \end{split} 
 \label{eq:weyl+g}
\end{align}
The solution of the second equation is given by 
\begin{align}
 g(y) =
 \begin{cases}
  \displaystyle -\frac{1}{2} \sinh^{-1} \left( \frac{\hbar v_F}{\xi E}
  \right) & (y < 0), \\
  \displaystyle +\frac{1}{2} \sinh^{-1} \left( \frac{\hbar v_F}{\xi E}
  \right) & (y > 0). 
 \end{cases}
 \label{eq:const-g(y)}
\end{align}
The sign of $g(y)$ changes across the zigzag edge, and 
this sign change indicates that the $z$-component of the pseudospin 
flips at the edge.
The flip is induced by the gauge field $A^{\rm q}_x(y)$.
To represent this, we integrate the first equation of Eq.~(\ref{eq:weyl+g})
from $y=-\xi_g$ to $\xi_g$, and acquire
\begin{align}
 -\int_{-\xi_g}^{\xi_g} \frac{dg(y)}{dy} dy
 = \frac{1}{\hbar} \int_{-\xi_g}^{\xi_g} A^{\rm q}_x(y) dy.
 \label{eq:w-s-integ}
\end{align}
We have neglected other terms, because they are proportional to
$\xi_g$ and become zero in the limit of $\xi_g =0$.
By substituting Eq.~(\ref{eq:const-g(y)}) into 
Eq.~(\ref{eq:w-s-integ}), we find
\begin{align}
 -\sinh^{-1} \left( \frac{\hbar v_F}{\xi E} \right) =
 \frac{1}{\hbar} \int_{-\xi_g}^{\xi_g} A^{\rm q}_x(y) dy.
 \label{eq:g-ene}
\end{align}
Hence, Eq.~(\ref{eq:const-g(y)}) becomes
\begin{align}
  g(y) =
 \begin{cases}
  \displaystyle + \frac{1}{2}\left(
  \frac{1}{\hbar} \int_{-\xi_g}^{\xi_g} A^{\rm q}_x(y) dy \right)
  & (y < 0), \\
  \displaystyle - \frac{1}{2}\left(
  \frac{1}{\hbar} \int_{-\xi_g}^{\xi_g} A^{\rm q}_x(y) dy \right) 
  & (y > 0). 
 \end{cases}
\label{eq:asol}
\end{align}
Having described the wave function of the localized state, 
let us now calculate $E$ and $\xi$.
To this end, we use the first equation of Eq.~(\ref{eq:weyl+g}) 
for $|y| \ge \xi_g$
and obtain 
\begin{align}
 \frac{E}{v_F}
 = \frac{p_x}{\cosh \left( \displaystyle 
 \frac{1}{\hbar} \int_{-\xi_g}^{\xi_g} A^{\rm q}_x(y) dy \right)}.
 \label{eq:ene}
\end{align}
Moreover, using Eq.(\ref{eq:g-ene}), we find
\begin{align}
 \frac{1}{\xi}
 = -k_x \tanh \left( \displaystyle 
 \frac{1}{\hbar} \int_{-\xi_g}^{\xi_g} A^{\rm q}_x(y) dy \right).  
 \label{eq:xi-n}
\end{align}
In addition to this localized state, 
there is another localized state for the same $k_x$ with 
the same $\xi$, but with the opposite sign of $E$.
This results from the particle-hole symmetry of the Hamiltonian,
and the wave function is given by $\sigma_z \Psi_{\rm K}(y)$.

In the following, we will show that 
the solutions can reproduce all the properties 
of the edge states known 
in the tight-binding lattice (TB) model,~\cite{tanaka87,fujita96,nakada96}
such as 
the asymmetric energy band structure with respect to the K (K$'$) point, 
the flat energy band, and the pseudospin structure.

The asymmetric energy band structure with respect to the K (K$'$) point 
originates from the normalization condition of the wave function,
which requires that $\xi$ should be positive.
This requirement restricts the value of $k_x$ in Eq.~(\ref{eq:xi-n}).
When $A^{\rm q}_x(y)$ is positive, 
Eq.~(\ref{eq:xi-n}) indicates that the localized states appear only 
at $k_x<0$ around the K point. 
This is the reason 
why the localized states appear in the energy spectrum
only at one side around the K point.
A similar argument can be used for the K$'$ point,
which concludes that the localized state 
appears at $k_x > 0$ around the K$'$ point.
The Hamiltonian around the K$'$ point is expressed by
\begin{align}
 H_{\rm K'}({\bf r})=
 v_F \bsigma' \cdot (\hat{\bf p} - {\bf A}^{\rm q}({\bf r})).
 \label{eq:weyl-K'}
\end{align}
Therefore, we obtain different signs in front of $\bf{A}^{\rm q}(\bf{r})$
in $H_{\rm K}(y)$ and Eq.~(\ref{eq:weyl-K'}), which 
causes the negative sign in front of the right-hand side of
Eq.~(\ref{eq:xi-n}) to disappear for the K$'$ point.
Thus, when $A^{\rm q}_x(y)$ is negative (Klein edges),
edge states appear on the opposite side; 
$k_x > 0$ around the K point and $k_x < 0$ around the K$'$ point.
Calculations on the TB model with Klein edges
also agree with the results obtained here.

A singularity of the gauge field, $|A^{\rm q}_x(y)| \to \infty$,
is the origin of the flat energy dispersion
and the pseudospin polarization of the edge states.
When $(1/\hbar) \int_{-\xi_g}^{\xi_g} A_x(y) dy \to \infty$, 
$E$ in Eq.~(\ref{eq:ene}) becomes zero.
The zero energy eigenvalue between the K and K$'$ points in the band
structure corresponds to the flat energy band of the edge
state.~\cite{fujita96}
Moreover, from Eq.~(\ref{eq:asol}), 
$g(y) \to \infty$ for $y < 0$ and $g(y) \to -\infty$ 
for $y > 0$ are obtained. In this case, 
the localized state is a pseudospin-up state
$\Psi_{\rm K}({\bf r}) \propto {}^t (1,0)$
for $y < 0$ and a pseudospin-down state
$\Psi_{\rm K}({\bf r}) \propto {}^t (0,1)$ for $y > 0$.
Hence, a singular gauge field at the zigzag edge causes polarization of 
the pseudospin of the localized states.
Polarization of the pseudospin means that 
the wave function has amplitude only at the A (or B) atom,
so that this result agrees with the result from the TB model for the
edge state.~\cite{fujita96}
Comparing Eqs.~(\ref{eq:xi-n}) and (\ref{eq:ene})
with Eqs.~(\ref{eq:kycond}) and (\ref{eq:solu1}), 
the relation between 
the variable $\phi$ and the field $A_x^{\rm q}(y)$ is observed as
$\phi = (1/\hbar)\int_{-\xi_g}^{\xi_g} A_x^{\rm q}(y)dy$.

Here, we note that $N$ in Eq.~(\ref{eq:phiR})
is not a function of $y$, but is a constant for $y \in (-\infty,\infty)$.
Therefore, the edge states appear
on both sides of the zigzag edge, $y>0$ and $y<0$, 
while the standing waves appear on only one side of the edge
under the limit $(1/\hbar)\int_{-\xi_g}^{\xi_g} A_x^{\rm q}(y)dy \to \pm \infty$.
With this limit, the edge states can be confined to one side of the edge,
because the energy of the localized state becomes $E = 0$,
and therefore, the superposition of an edge state, $\Psi_{\rm K}(y)$,
and its electron-hole pair state, $\sigma_z \Psi_{\rm K}(y)$,
is a solution.
It is easy to see that $\Psi_{\rm K}(y) + \sigma_z \Psi_{\rm K}(y)$ 
has amplitude only for $y <0$, while 
$\Psi_{\rm K}(y) - \sigma_z \Phi_{\rm K}(y)$ 
has amplitude only for $y >0$.
The wave function of the edge state for $y<0$ is then given by
\begin{align}
 \Psi_{{\rm K},k_x<0}({\bf r}) 
 = \frac{e^{ik_x x}}{\sqrt{L_x}}\sqrt{2|k_x|} e^{k_x |y|}
 \begin{pmatrix}
  1 \cr 0
 \end{pmatrix},
 \label{eq:eswavefunc}
\end{align}
where the normalization constant has been fixed,
$\sqrt{2|k_x|}$, 
by assuming that the system is a semi-infinite graphene plane.
We note that the mass term, $m \sigma_z$, is proportional to the 
particle-hole symmetry operator, $\sigma_z$. 
Thus, the mass term automatically restricts the region where
the edge states can appear ($y>0$ or $y<0$), and this
is shown in Appendix~\ref{app:edge}.

Finally, we consider the edge states in nanoribbons.
Note first that the exact localization length for the case of 
a zigzag nanoribbon with width $L$ satisfies
\begin{align}
\begin{split}
 & \frac{1}{\xi} = -k_x \tanh\left(\frac{L}{\xi}\right), \\
 & \left[ \leftrightarrow -k_x L = (k_x \xi) {\rm atanh} \left(\frac{1}{k_x \xi }\right)\right],
\end{split}
 \label{eq:xiexact}
\end{align}
which is obtained by analytical continuation $k_y = i/\xi$ for
Eq.~(\ref{eq:Bery}). 
Comparing this equation with Eq.~(\ref{eq:xi-n}) shows 
that the large value of the gauge field in Eq.~(\ref{eq:xi-n})
corresponds to the case of $L/\xi \gg 1$ in Eq.~(\ref{eq:xiexact}).
This is consistent with having solved the Hamiltonian locally near the edge, 
in which it was implicitly assumed that the condition $L/\xi \gg 1$ is satisfied.
Except the edge states whose localization length is in the order of $L$, 
Eqs.~(\ref{eq:xi-n}) and (\ref{eq:xiexact}) give almost identical
values of $\xi \simeq -k^{-1}_x$, 
which justifies the description using the gauge field.
Note also that the condition $L/\xi \gg 1$ also represents the condition
$k_x L \ll -1$, which is clear from the second equation in
Eq.~(\ref{eq:xiexact}). 
To solve the Hamiltonian for the edge states with $\xi = {\cal O}(L)$, 
Eq.~(\ref{eq:weyl+g}) must be solved globally, 
for example, on a circle, 
which is a challenging issue.

\subsection{Mass Term}\label{ssec:mass}

Let us reconsider the scattering problem for the case where 
the Hamiltonian includes a mass term. 
The total Hamiltonian is given by 
$H^{m}_{\rm K}(y) \equiv H_{\rm K}(y) + m \sigma_z$,
where the mass, $m$, is a constant over the space $y \in (-\infty,+\infty)$. 
The solutions of $H^{m}_{\rm K}(y)$ can be constructed
from the solutions of $H_{\rm K}(y)$ as follows.

For $H_{\rm K}(y)$,
the standing wave solutions, 
Eqs.~(\ref{eq:s-sol}) and (\ref{eq:s-solv}), satisfy
\begin{align}
\begin{split}
 & H_{\rm K}(y) \Psi^{c}_{{\rm K},{\bf k}}(y) = 
 \hbar v_{\rm F}k \Psi^{c}_{{\rm K},{\bf k}}(y), \\
 & H_{\rm K}(y) \Psi^{v}_{{\rm K},{\bf k}}(y) = 
 - \hbar v_{\rm F}k
 \Psi^{v}_{{\rm K},{\bf k}}(y).
\end{split}
\end{align}
For the mass term, 
because $\Psi^{v}_{{\rm K},{\bf k}}(y)=\sigma_z \Psi^{c}_{{\rm K},{\bf k}}(y)$,
we obtain
\begin{align}
\begin{split}
 & m\sigma_z \Psi^{c}_{{\rm K},{\bf k}}(y) 
 = m \Psi^{v}_{{\rm K},{\bf k}}(y), \\
 & m\sigma_z \Psi^{v}_{{\rm K},{\bf k}}(y)
 = m \Psi^{c}_{{\rm K},{\bf k}}(y).
\end{split}
\end{align}
Thus, by changing the basis state from 
$|\Psi^{c}_{{\rm K},{\bf k}}\rangle$ and 
$|\Psi^{v}_{{\rm K},{\bf k}}\rangle$
into $|\Psi_{{\rm K},{\bf k},{\rm A}({\rm B})}\rangle$
$[\equiv (1/\sqrt{2})(|\Psi^{c}_{{\rm K},{\bf
k}}\rangle\pm|\Psi^{v}_{{\rm K},{\bf k}}\rangle)]$, 
the Hamiltonian is represented as 
\begin{align}
 H^{m}_{\rm K} 
 &\to
\begin{pmatrix}
 \langle \Psi_{{\rm K},{\bf k},{\rm A}}|H^{m}_{\rm K}| \Psi_{{\rm
 K},{\bf k},{\rm A}} \rangle & 
 \langle \Psi_{{\rm K},{\bf k},{\rm A}}|H^{m}_{\rm K}| \Psi_{{\rm
 K},{\bf k},{\rm B}} \rangle\cr
\langle \Psi_{{\rm K},{\bf k},{\rm B}}|H^{m}_{\rm K}| \Psi_{{\rm K},{\bf
 k},{\rm A}} \rangle &
 \langle \Psi_{{\rm K},{\bf k},{\rm B}}|H^{m}_{\rm K}| \Psi_{{\rm
 K},{\bf k},{\rm B}} \rangle
\end{pmatrix} \nn \\
 &=
 \begin{pmatrix}
 m & \hbar v_{\rm F}k \cr 
 \hbar v_{\rm F}k & -m
 \end{pmatrix}.
 \label{eq:HKmat}
\end{align}
Here, the angle $\phi_k$ is defined as
\begin{align}
 \cos \phi_k \equiv \frac{m}{E_k}, \ \
 \sin \phi_k \equiv \frac{\hbar v_{\rm F}k}{E_k}, 
 \label{eq:defphi}
\end{align}
where $E_k \equiv \sqrt{m^2+(\hbar v_{\rm F}k)^2}$.
The normalized eigenvectors 
of the matrix in Eq.~(\ref{eq:HKmat}) are then
\begin{align}
\begin{pmatrix}
 \cos\frac{\phi_k}{2} \cr 
 \sin\frac{\phi_k}{2}
\end{pmatrix} \ \ {\rm and} \ \ 
\begin{pmatrix}
 -\sin \frac{\phi_k}{2} \cr 
 \cos \frac{\phi_k}{2}
\end{pmatrix} 
\label{eq:bfieldp}
\end{align}
for the $E_k$ and $-E_k$ eigenvalues, respectively.
$\Psi_{{\rm K},{\bf k},{\rm A}}(y)=
\sqrt{2}N(y)\sin(k_y y -\theta({\bf k}))$ and
$\Psi_{{\rm K},{\bf k},{\rm B}}(y)=\sqrt{2}N(y)\sin(k_y y)$ 
are obtained from Eqs.~(\ref{eq:s-sol}) and (\ref{eq:s-solv}); therefore, 
the standing wave near the zigzag edge is given by
\begin{align}
\begin{split}
 & \Psi^{m,c}_{{\rm K},{\bf k}}(y) = \sqrt{2}N(y) 
 \begin{pmatrix}
  \cos \left(\frac{\phi_k}{2}\right) \sin \left( k_y y - \theta({\bf k}) \right)\cr 
  \sin \left(\frac{\phi_k}{2} \right) \sin \left( k_y y \right)
 \end{pmatrix},
 \\
 & \Psi^{m,v}_{{\rm K},{\bf k}}(y) = \sqrt{2}N(y)
 \begin{pmatrix}
 - \sin \left( \frac{\phi_k}{2}\right) \sin \left( k_y y - \theta({\bf k}) \right)\cr 
  \cos \left( \frac{\phi_k}{2}\right) \sin \left( k_y y \right)
 \end{pmatrix},
\end{split}
\label{eq:mass-wave}
\end{align}
where we have omitted to write the plane wave parallel to the edge. 
The factors $\cos \left(\phi_k/2\right)$
and $\sin \left(\phi_k/2 \right)$ appear in a manner 
similar to the eigenvalue problem of the spin magnetic moment
in a magnetic field.

\subsection{External Magnetic Field}\label{ssec:emf}

In this subsection, 
solutions are constructed for a magnetic field 
applied perpendicular to the graphene plane.~\cite{mcclure56}
A magnetic field $B$ can be represented by 
the electromagnetic gauge field as ${\bf A}(y)=(By,0)$.
This gauge field is included in the Hamiltonian 
by substituting the momentum operator ${\bf {\hat p}}$ with
${\bf {\hat p}}-e{\bf A}$.
For the case $A^{\rm q}_x(y)=0$, 
the eigenequation becomes
\begin{align}
 v_{\rm F} \left[
 \sigma_x (\hat{p}_x-eBy) + \sigma_y \hat{p}_y
 \right] \Phi({\bf r}) = E \Phi({\bf r}),
\end{align}
and the solutions are given by the Landau states,
which are specified by an integer $n$ and 
a center coordinate $Y$ as
\begin{align}
 \Phi^{\rm LL}_{nY}({\bf r}) 
 &= C_{nY} e^{i\frac{Yx}{{\it l}^2}}
 e^{-\frac{1}{2}\left(\frac{y-Y}{\it l}\right)^2} \nn \\
 & \times 
 \begin{pmatrix}
  {\rm sgn}(n)\sqrt{2|n|}H_{|n|-1}\left((y-Y)/{\it l}\right) \cr
  -H_{|n|}\left((y-Y)/{\it l}\right)
 \end{pmatrix},
 \label{eq:LL}
\end{align}
where $C_{nY}$ is a normalization constant,
${\it l} = \sqrt{\hbar/eB}$, and $H_n(x)$ is a Hermite
polynomial defined by 
$H_n(x) \equiv (-1)^n e^{x^2} (d/dx)^n e^{-x^2}$ ($n \ge 0$).
The energy eigenvalue of $\Phi^{\rm LL}_{nY}({\bf r})$ is given by
$E_n = {\rm sgn}(n) \sqrt{2|n|}\hbar v_{\rm F}/{\it l}$.

A method similar to that in Sec.~\ref{sec:exs} is used
to solve the scattering problem in the presence of a magnetic field.
The energy eigenstate of $H_{\rm K}({\bf r})$ is parameterized as
$\Psi_{\rm K}({\bf r}) = N(y) \Phi^{\rm LL}_{nY}({\bf r})$.
Substituting this into 
$H_{\rm K}({\bf r})\Psi_{\rm K}({\bf r})=E_n\Psi_{\rm K}({\bf r})$,
and using 
$H^0_{\rm K}({\bf r})\Phi^{\rm LL}_{nY}({\bf r})=E_n\Phi^{\rm LL}_{nY}({\bf r})$,
we obtain the constraint equation
for $N(y)$ and $\Phi^{\rm LL}_{nY}({\bf r})$,
\begin{align}
 \left\{ \sigma_y [\hat{p}_y N(y)] + \sigma_x A_x^{\rm q}(y)N(y) \right\}
 \Phi^{\rm LL}_{nY}({\bf r}) =0.
\end{align}
Two cases can be considered as a solution 
for this successive equation
[see Eqs.~(\ref{eq:case1}) and (\ref{eq:case2})].
Here, we choose the case where
\begin{align}
\begin{split}
 & A_x^{\rm q}(y)N(y) + \hbar \frac{dN(y)}{dy} = 0, \\
 & \Phi^{\rm LL}_{nY,{\rm B}}({\bf r}) = 0,
\end{split} \ \ \
 (|y|\le \xi_g).
\end{align}
From Eq.~(\ref{eq:LL}),
the second equation leads to
\begin{align}
 C_{nY}e^{-\frac{1}{2}\left(\frac{Y}{\it l}\right)^2}
 H_{|n|}\left(-2Y/{\it l}\right) = 0,
 \label{eq:constB}
\end{align}
with the limit $\xi_g \to 0$ ($y\to 0$).
The number of zeros of $H_n$ is $|n|$,
so that there are $|n|$ solutions of Eq.~(\ref{eq:constB}),
which are denoted as 
$Y_i$ ($i=0,\cdots,|n|$).
The solutions can then be written as
\begin{align}
 \Psi_{{\rm K},nY_i}({\bf r}) = N(y) \Phi^{\rm LL}_{nY_i}({\bf r}).
 \label{eq:LLz}
\end{align}
Note that $\Psi_{{\rm K},nY}({\bf r})$ 
with a large value of $Y$ that satisfies $Y \gg {\it l}$, 
can be an approximate solution, 
due to the exponential factor in Eq.~(\ref{eq:constB}).
The solution with a large value of
$Y$ represents the wave function in the bulk, 
and is not sensitive to the details of the edge.
The solutions given in Eq.~(\ref{eq:LLz})
concern the Landau states near the zigzag edge, 
and these are examined in the following.

For the case that $n$ is an odd integer, 
$Y=0$ satisfies Eq.~(\ref{eq:constB}),
because $H_n(0)=0$.
The wave function with $Y=0$ decays according to 
$\exp(-y^2/2{\it l}^2)$, and 
the amplitude has a maximum at the zigzag edge.
Note that the localization length is in the order of ${\it l}$
(${\it l}\simeq 25[{\rm nm}]/\sqrt{B[{\rm T}]}$),
which is larger than the localization length of the edge state
$\xi=-k_x^{-1}$ where $k_x^{-1}$ takes a value of the same order as the
lattice constant.

The lowest Landau level ($n=0$) 
can not satisfy the condition of Eq.~(\ref{eq:constB})
because $H_0(-2Y/{\it l})=1$ and 
the amplitudes of B-atoms do not vanish at the edge.
Thus, the lowest Landau level is absent for the K point.
On the other hand, 
the lowest Landau level appears for the K$'$ point. 
The Hamiltonian for the K$'$ point is given by
\begin{align}
 H_{\rm K'}({\bf r})=v_{\rm F}
 \left[
 -\sigma_x (\hat{p}_x -eA_x(y) - A^{\rm q}_x(y)) + \sigma_y \hat{p}_y
 \right].
\end{align}
For the case of $A^{\rm q}_x(y)=0$,
$H_{\rm K}({\bf r})$ and $H_{\rm K'}({\bf r})$ 
are related as
$H_{\rm K'}({\bf r})=\sigma_y H_{\rm K}({\bf r})\sigma_y$,
and therefore the solutions for the K$'$ point 
are given by $\sigma_y \Phi^{\rm LL}_{nY'}({\bf r})$.
The constraint equation for the K$'$ point is 
\begin{align}
 \left\{ \sigma_y [\hat{p}_y N(y)] + \sigma_x A_x^{\rm q}(y)N(y) \right\}
 \sigma_y \Phi^{\rm LL}_{nY'}({\bf r}) =0,
\end{align}
which reduces to the condition 
$\Phi^{\rm LL}_{nY',{\rm A}}(y) = 0$.
The solution is then given by
\begin{align}
 \Psi_{{\rm K'},nY'_j}({\bf r}) = N(y) \sigma_y 
 \Phi^{\rm LL}_{nY'_j}({\bf r}), 
 \label{eq:LLK'}
\end{align}
where $Y'_j$ denotes the solution of the constraint equation,
\begin{align}
 C_{nY} e^{-\frac{1}{2}\left(\frac{Y}{{\it l}}\right)^2} \sqrt{2|n|}
 H_{|n|-1}\left(-Y/{\it l}\right)=0.
\end{align}
This condition is satisfied for $n=0$, 
so that the lowest Landau level appears for the K$'$ point. 
There is no constraint for the value of $Y$.
For the case of the first Landau levels ($n=\pm 1$), 
the Landau level for the K point appears, while 
that for the K$'$ point disappears. 
Therefore, near the zigzag edge,
the lowest and first Landau levels 
are not symmetric with respect to the K and K$'$ points.

\subsection{Local Density of States}\label{ssec:ldos}

Several groups have conducted scanning tunneling spectroscopy (STS)
measurements to determine the LDOS near the step
edge of graphite.~\cite{klusek00,giunta01,kobayashi05,niimi05} 
A peak structure in the LDOS due to the edge states 
has been extensively discussed by many authors. Here, we calculate the LDOS near the zigzag edge.
We show that some characteristic features that originate from the
pseudospin polarization, the edge states, and the mass appear in the
LDOS.

Let us first review the LDOS for 
graphene without an edge. 
Assuming that electrons are non-interacting,
the bulk LDOS is given by 
\begin{align}
 \rho(E) 
 = \frac{1}{2\pi} \frac{|E|}{(\hbar v_{\rm F})^2},
 \label{eq:ldos_bulk}
\end{align}
where $\rho(E)$ is proportional to $|E|$, 
which results from the Dirac cone spectrum.
Note that the actual LDOS is given by $g_s g_v \rho(E)$,
where $g_s=2$ ($g_v=2$) accounts for the spin (valley) degrees of freedom. 
Next, the LDOS near the zigzag edge
is calculated using the solutions given in Eq.~(\ref{eq:mass-wave}).
The LDOS has the form,
\begin{align}
 \rho_s(E,y)= \frac{1}{2\pi} \frac{|E|}{(\hbar v_{\rm F})^2} R(E,y),
\end{align}
where $R(E,y)$ is defined as
\begin{align}
 R(E,y) \equiv
 \frac{1}{\pi}\int^\pi_0 d\theta 
 \Psi^m_{{\rm K},{\bf k}}(y)^\dagger
 \Psi^m_{{\rm K},{\bf k}}(y).
 \label{eq:Rfac}
\end{align}
By performing the integral with respect to the angle $\theta$ 
in Eq.~(\ref{eq:Rfac}), we obtain an analytical result for $R(E,y)$ as
\begin{align}
 R(E,y) =
\begin{cases}
 \displaystyle F\left(k|y|\right) + \frac{m}{|E|} G\left(k|y|\right) \ \ (E>0), \\
 \displaystyle F\left(k|y|\right) - \frac{m}{|E|} G\left(k|y|\right) \ \ (E<0),
\end{cases}
\label{eq:ldos}
\end{align}
where $k$ is a function of $E$ 
according to $k=\sqrt{E^2 - m^2}/(\hbar v_{\rm F})$,
and the functions $F$ and $G$ are defined as
\begin{align}
\begin{split}
 & F(k|y|) \equiv 
 1-\left\{ \frac{J_0(2k|y|)+J_2(2k|y|)}{2} \right\}, \\
 & G(k|y|) \equiv \frac{J_0(2k|y|)-J_2(2k|y|)}{2}.
\end{split} 
\end{align}
Here, $J_\nu(x)$ is a Bessel function of order $\nu$.

\begin{figure}[htbp]
 \begin{center}
  \includegraphics[scale=0.55]{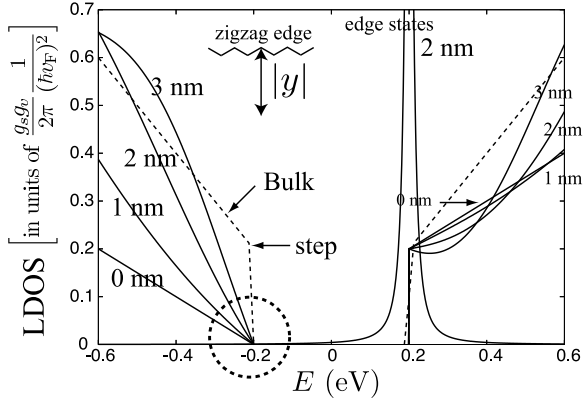}
 \end{center}
 \caption{
 Positional dependence of the LDOS structure for the cases of
 $m=0.2$[eV].
 The number located on each solid line represents the distance 
 (corresponding to $|y|$ in the inset) from the zigzag edge. 
 The LDOS at $E=-m$ vanishes near the zigzag edge, 
 which is emphasized by the dashed circle.
 The dashed line denotes the LDOS in the bulk which is defined by the
 LDOS at $|y|\to \infty$. 
 A peak structure due to the edge states is plotted for comparison. 
 Note that there are several intrinsic
 perturbations~\cite{sasaki09-hd} 
 that can change the position of the peak.
 }
 \label{fig:ldos}
\end{figure}

Because the case of $m=0$ has been considered elsewhere,~\cite{sasaki10-forward}
we consider the case $m\ne 0$ here.
Eq.~(\ref{eq:ldos_bulk}) holds for $|E| \ge |m|$. 
The bulk LDOS vanishes for the case $|E| < |m|$, 
as shown by the dashed line in Fig.~\ref{fig:ldos}.
Note that the LDOS disappears suddenly at $E=\pm m$, 
and the bulk LDOS has a step like structure at $E=\pm |m|$.
The bulk LDOS is symmetric with respect to $E=0$, 
even for the case $m \ne 0$.
However, note that the LDOS near the edge 
is not symmetric for the case of $m \ne 0$,
which is clear from the different signs in front of 
the function $G$ in Eq.~(\ref{eq:ldos}).
In Fig.~\ref{fig:ldos},
the LDOS are plotted at $|y|=0$, 1, 2, and 3 [nm]
for the case of $m=0.2$ eV. 
Note that for the case $m=-0.2$ eV,
the corresponding LDOS curve 
is given by interchanging the conduction and valence bands in
Fig.~\ref{fig:ldos}.

The asymmetry in the LDOS near the edge 
appears at the following points.
First, a step structure appears only at $E=0.2$ eV.
At $E=-0.2$ eV, the LDOS vanishes, and the step 
structure is absent, as indicated by the dashed circle in
Fig.~\ref{fig:ldos}.
The absence of the LDOS at $E=-0.2$ eV 
can be explained by the zigzag edge consisting of A-atoms 
makes the standing wave 
polarized into A-atoms near the zigzag edge.
However, eigenstates with energy $E=-m$ should be polarized 
into B-atoms by the factors in Eq.~(\ref{eq:bfieldp}),
and the amplitude of A-atoms are strongly suppressed
by the mass term. 
Therefore, electrons with energy $E=-m$
can not approach the zigzag edge, and therefore the LDOS
disappears.
Secondly, the LDOS peak of the edge states
appears only at $E=0.2$ eV.
This is a straightforward consequence of
the edge state amplitude appearing only for A-atoms.
The absence of the LDOS at $E=-m$ 
and the presence of the peak at $E=m$ due to the edge states 
occurs at different sides of the band edge.
To plot the LDOS of the edge states in Fig.~\ref{fig:ldos}, 
we have used 
\begin{align}
 \rho_e(E,y) = \frac{1}{2\pi^2}
 \frac{2\delta}{(E-m)^2+\delta^2}\frac{1}{4y^2},
 \label{eq:dosed}
\end{align}
where $\delta$ is a phenomenological parameter that 
represents the energy uncertainty of the edge states, 
for which we assume $\delta = 10$ meV.
This result has been derived in Ref.~\onlinecite{sasaki10-forward}
for the case of $m=0$.
Note that $\rho_e(E,y)$ decreases as $\sim y^{-2}$, 
which is a slowly decreasing function 
compared with the exponential decay wave function of the edge state.

\section{Armchair edge}\label{sec:arm}

In this section, 
the scattering problem for the armchair edge is solved
using a method similar to that used in Sec.~\ref{sec:zig}.
The standing wave shows that 
the pseudospin does not change its direction 
through the reflection at the armchair edge.

\subsection{Standing Waves}

Solutions for the case of $\phi^{\rm q}(x)=0$ 
in Eq.~(\ref{eq:armchairH}) are constructed first, 
and then used as the basis functions to construct the standing wave 
near the armchair edge.
Let $\Phi(x)$ represent the solution of the unperturbed Hamiltonian,
$H_0(x)=v_{\rm F}(\tau_3 \sigma_x \hat{p}_x + \tau_0 \sigma_y p_y)$.
The perturbed Hamiltonian satisfies 
$H(-x) = \tau_1 H(x) \tau_1$,
so that the functions $\Phi(x)$ that satisfy the constraint equation
\begin{align}
 \tau_1 \Phi(-x) = e^{-ig} \Phi(x), \ \ 
 (g=0,\pi)
 \label{eq:xmirror}
\end{align}
are useful for construction of solutions in the case of $\phi^{\rm q}(x)\ne 0$.
From Eq.~(\ref{eq:xmirror}), 
we may write 
\begin{align}
 \Phi(x) =
 \begin{pmatrix}
  \Phi_{\rm K}(x) \cr e^{ig} \Phi_{\rm K}(-x)
 \end{pmatrix}.
 \label{eq:wf_arm}
\end{align}
By using Eq.~(\ref{eq:wf_arm}),
the energy eigenequation becomes
\begin{align}
\begin{split}
 & \left( \frac{E}{\hbar v_{\rm F}}-\sigma_y k_y \right) \Phi_s(x) 
 = -i \sigma_x \frac{d}{dx} \Phi_a(x), \\
 & \left( \frac{E}{\hbar v_{\rm F}}-\sigma_y k_y \right) \Phi_a(x) 
 = -i \sigma_x \frac{d}{dx} \Phi_s(x).
\end{split}
 \label{eq:dirac3}
\end{align}
where $\Phi_s(x)$ and $\Phi_a(x)$ are defined as
\begin{align}
\begin{split}
 & \Phi_s(x) \equiv 
 e^{-i\frac{g}{2}} \Phi_{\rm K}(x)+e^{+i\frac{g}{2}} \Phi_{\rm K}(-x),
 \\
 & \Phi_a(x) \equiv 
 e^{-i\frac{g}{2}} \Phi_{\rm K}(x)-e^{+i\frac{g}{2}} \Phi_{\rm K}(-x).
\end{split}
 \label{eq:psisa_def}
\end{align}
For the case $g=0$,
$\Phi_s(x)$ is an even function,
while $\Phi_a(x)$ is an odd function.
For the case of $g=\pi$, 
$\Phi_s(x)$ is an odd function, 
while $\Phi_a(x)$ is an even function.

For the case $g=0$, we can set
\begin{align}
\begin{split}
 & \Phi_s(x) = \cos (k_x x) \phi, \\
 & \Phi_a(x) = i \sin (k_x x) \phi.
\end{split}
\label{eq:cond_sa-1}
\end{align}
Substituting these into Eq.~(\ref{eq:dirac3}),
we obtain the secular equation:
\begin{align}
 \left( \frac{E}{\hbar v_{\rm F}}-\bsigma\cdot {\bf k} \right) \phi =0.
\end{align}
The solutions of this secular equation satisfy 
$E=\pm \hbar v_{\rm F}k$, and
the eigenfunction in the conduction band is given by
$\phi_{{\rm K},{\bf k}}^c$, which is defined as
\begin{align}
 \phi_{{\rm K},{\bf k}}^c = \frac{1}{\sqrt{2}}
 \begin{pmatrix}
  e^{-i\theta({\bf k})} \cr 1
 \end{pmatrix}.
 \label{eq:blochf}
\end{align}
By substituting Eq.~(\ref{eq:cond_sa-1}) into Eq.~(\ref{eq:psisa_def}),
we obtain $\Phi_{\rm K}(x)=e^{ik_x x} \phi_{{\rm K},{\bf k}}^c /2$. 
Using Eq.~(\ref{eq:wf_arm}), it can be seen that 
\begin{align}
 \Phi^{0}(x) = \phi_{{\rm K},{\bf k}}^c
 \begin{pmatrix}
  e^{+ik_x x} \cr e^{-ik_x x} 
 \end{pmatrix}.
 \label{eq:aphi0}
\end{align}
Similarly, for the case $g=\pi$, we have
\begin{align}
 \Phi^{\pi}(x) =\phi_{{\rm K},{\bf k}}^c
 \begin{pmatrix}
  e^{+ik_x x} \cr -e^{-ik_x x} 
 \end{pmatrix}.
 \label{eq:aphipi}
\end{align}
New basis functions are defined using Eqs.~(\ref{eq:aphi0}) and (\ref{eq:aphipi}),
as
\begin{align}
\begin{split}
 & \Phi^{\rm K}(x) \equiv \frac{1}{2} \left(
 \Phi^{0}(x)+\Phi^{\pi}(x)\right) = 
\phi_{{\rm K},{\bf k}}^c
 \begin{pmatrix}
  e^{+ik_x x} \cr 0
 \end{pmatrix}, \\
 & \Phi^{\rm K'}(x) \equiv \frac{1}{2} \left(
 \Phi^{0}(x)-\Phi^{\pi}(x)\right) = 
\phi_{{\rm K},{\bf k}}^c
 \begin{pmatrix}
  0 \cr e^{-ik_x x}
 \end{pmatrix}.
\end{split}
\end{align}
The eigenstate $\Phi^{\rm K}(x)$ represents
a free propagating state with momentum ${\bf k}$ near the K point, 
while $\Phi^{\rm K'}(x)$ represents a state with momentum ${\bf k'}=(-k_x,k_y)$
near the K$'$ point.
It is clear that these are eigenstates 
in the absence of the edge.
In the presence of the armchair edge,
neither $\Phi^{\rm K}(x)$ nor $\Phi^{\rm K'}(x)$ is an eigenstate,
but a true eigenstate is the standing wave 
that is given by a superposition between 
$\Phi^{\rm K}(x)$ and $\Phi^{\rm K'}(x)$ as
\begin{align}
 \Psi(x) = c^{\rm K}(x)\Phi^{\rm K}(x)+c^{\rm K'}(x)\Phi^{\rm K'}(x).
\end{align}
To find $c^{{\rm K},{\rm K'}}(x)$,
it is useful to rewrite the total Hamiltonian as
\begin{align}
 H(x) = H_0(x) + v_{\rm F} \sigma_x \left[ 
 \phi^{\rm q}_{\it r}(x) \tau_1 + \phi^{\rm q}_{\it i}(x) \tau_2
 \right],
 \label{eq:Hdec}
\end{align}
where $\phi^{\rm q}(x)$
is expressed in terms of real and imaginary parts, as 
$\phi^{\rm q}(x) \equiv \phi^{\rm q}_{\it r}(x)-i \phi^{\rm q}_{\it i}(x)$.
In Sec.~\ref{sec:model}, we have shown that 
$\phi^{\rm q}(x)\equiv A^{\rm q}_x(x) e^{2i[\varphi(x)-k_{\rm F}x]}$,
where $A_x^{\rm q}(x) = \partial_x \varphi(x)$.
$A^{\rm q}_x(x)$ is an even function with respect to $x$; therefore, 
$\varphi(x)$ can be taken as an odd function,
so that the field satisfies $\phi^{\rm q}(-x)=\phi^{\rm q}(x)^*$.
From this condition, it follows that
$\phi^{\rm q}_{\it r}(x)$ is an even function,
while $\phi^{\rm q}_{\it i}(x)$ is an odd function.

Next, we construct solutions for the case $\phi^{\rm q}_{\it r}(x)=0$.
Let us define $\varphi_1(x)$ and $\varphi_2(x)$ using a real function $f(x)$ as 
\begin{align}
 \begin{pmatrix}
  \varphi_1(x) \cr
  \varphi_2(x)
 \end{pmatrix}
 = 
 \begin{pmatrix}
  \cosh f(x) & \sinh f(x) \cr
  \sinh f(x) & \cosh f(x)
 \end{pmatrix}
 \begin{pmatrix}
  \Phi^{\rm K}(x) \cr \Phi^{\rm K'}(x)
 \end{pmatrix}.
\end{align}
Since $\Phi^{\rm K}(x)$ and $\Phi^{\rm K'}(x)$ 
are the solutions of $H_0(x)$, 
we obtain the following equations for $f(x)$ from
$H(x)\varphi_i(x)= E\varphi_i(x)$,
\begin{align}
 \begin{pmatrix}
  \hat{p}_x & -i\phi_{\it i}(x) \cr
  i\phi_{\it i}(x) & -\hat{p}_x
 \end{pmatrix}
 \begin{pmatrix}
  \cosh f(x) & \sinh f(x) \cr
  \sinh f(x) & \cosh f(x)
 \end{pmatrix}
 =0.
\end{align}
All four components of this matrix are reduced into 
the same differential equation: 
$\partial_x f(x) = -\phi_{\it i}(x)/\hbar$.
$\phi_{\it i}(x)$ is an odd function,
so that we have $f(\xi_g)=f(-\xi_g)$
by using $\int_{-\xi_g}^{\xi_g} \phi_{\it i}(x)dx = 0$.
Because $f(x)=0$ when $\phi_{\it i}(x)=0$,
the constant of integration can be taken as zero.
As a result, we have $f(\xi_g)=f(-\xi_g)=0$.
Therefore, $f(x)$ can take only a non zero value 
for $|x|\le \xi_g$,
and the mixing between $\Phi^{\rm K}(x)$ and $\Phi^{\rm K'}(x)$
is negligible in the bulk.

\begin{figure}[htbp]
 \begin{center}
  \includegraphics[scale=0.5]{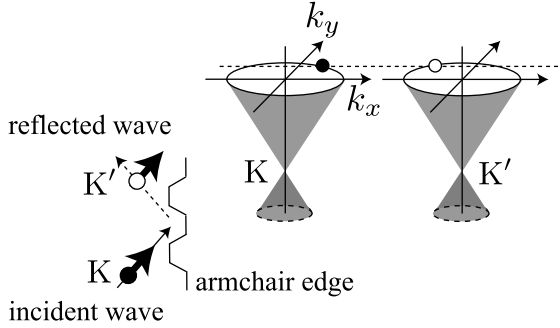}
 \end{center}
 \caption{
 The armchair edge reflects the wave vector 
 ${\bf k}=(k_x,k_y)$ of one valley 
 into ${\bf k'}=(-k_x,k_y)$ of another valley, and
 the two wave functions of the different valleys form a standing wave.
 The pseudospin is unchanged by the armchair edge.
 Note that the pseudospin for states near the K$'$ point 
 is not parallel to the vector ${\bf k'}$, while 
 the pseudospin for states near the K point 
 is parallel to the vector ${\bf k}$.
 }
 \label{fig:pseudospin-arm}
\end{figure}

Finally, we assume that 
the solution of the total Hamiltonian of Eq.~(\ref{eq:Hdec}) 
has the form of 
\begin{align}
 \Psi_\pm(x) = N(x) \left[ \varphi_1(x) \mp i \varphi_2(x) \right].
\end{align}
The constraint equation for $N(x)$ is then given by
\begin{align}
 \phi_{\it r}(x) N(x) \pm \hbar \frac{dN(x)}{dx} = 0.
 \label{eq:difN}
\end{align}
This constraint equation has the same form 
as Eqs.~(\ref{eq:case1}) and
(\ref{eq:case2}). 
Performing the integral for $x$ from $-\xi_g$ to $\xi_g$
in Eq.~(\ref{eq:difN}) gives 
\begin{align}
 \frac{N(+\xi_g)}{N(-\xi_g)}= 
 \exp \left( \mp \frac{1}{\hbar} \int_{-\xi_g}^{\xi_g}\phi_{\it r}(x)dx \right).
\end{align}
As we have shown in Fig.~\ref{fig:graphene}, 
$\phi_{\it r}(x)$ is a negative large quantity.
Thus, $\Psi_+(x)$ has an amplitude only for $x>\xi_g$,
while $\Psi_-(x)$ has an amplitude only for $x<-\xi_g$.
$\varphi_1(x) = \Phi^{\rm K}(x)$ and $\varphi_2(x) = \Phi^{\rm K'}(x)$
for $|x|\ge \xi_g$; therefore, 
the standing wave near the armchair edge 
is written as
\begin{align}
 \Psi^c_{{\bf k}}({\bf r}) 
 = \frac{e^{ik_y y}}{\sqrt{L_y}}N(x) \phi^c_{{\rm K},{\bf k}}
  \begin{pmatrix}
  e^{+ik_x x} \cr \mp i e^{-ik_x x} 
  \end{pmatrix},
\label{eq:armwf}
\end{align}
Note that the Bloch functions for the K and K$'$ points
are the same, which indicates that
the pseudospins of the incident and reflected waves are equal,
as shown in Fig.~\ref{fig:pseudospin-arm}.
Thus, the Berry's phase of the standing wave near the armchair edge is
given by $-\pi$,
which is in contrast to the case of the zigzag edge.~\cite{sasaki10-forward}
The boundary condition for the armchair edge
does not forbid an electronic state 
to cross the Dirac singularity point, 
and therefore the electron can pick up 
a nontrivial Berry's phase.

To understand the behavior of the standing wave in more detail, 
the density of $\tau_\alpha$ was examined.
The density for an eigenstate $\Psi(y)$ is defined by
the expected value of $\tau_\alpha$ as 
$\tau_\alpha(x) \equiv \Psi^\dagger(x) \tau_\alpha \Psi(x)$.
It is then straightforward to check from Eq.~(\ref{eq:armwf}) that
$\tau_1(x) \propto \pm \sin(2k_x x)$,
$\tau_2(x) \propto \pm \cos(2k_x x)$, and
$\tau_3(x) =0$.
$\tau_1(x)$ vanishes near the armchair edge (at $x=0$), and 
$\tau_2(x)$ takes a maximum value at the edge. 
This behavior can be understood from Eq.~(\ref{eq:Hdec}), 
in which $\tau_1$ couples with $\phi_{\it r}(x)$.
Since $\phi_{\it r}(x)$ is singular at $x=0$,
$\tau_1(x)$ can not have a non-zero value at $x=0$.
The result $\tau_3(x)=0$ indicates that
time-reversal symmetry is preserved.

\subsection{External Magnetic Field}

Let us examine the Landau states near the armchair edge.
The electromagnetic gauge field ${\bf A}= (0,-Bx)$ 
for an external magnetic field $B$
is included in the Hamiltonian $H({\bf r})$ of Eq.~(\ref{eq:H})
by the substitution ${\bf {\hat p}} \to {\bf {\hat p}}-e{\bf A}$.
The Hamiltonian satisfies 
$H({\bf -r}) = \tau_1 \sigma_x H({\bf r}) \sigma_x \tau_1$,
and therefore the solution can be written as
\begin{align}
 \Psi({\bf r}) =
 \begin{pmatrix}
  \Psi_{\rm K}({\bf r}) \cr \Psi_{\rm K'}({\bf r})
 \end{pmatrix}
 =
 \begin{pmatrix}
  \Psi_{\rm K}({\bf r}) \cr e^{ig} \sigma_x \Psi_{\rm K}({\bf -r})
 \end{pmatrix}.
 \label{eq:armcy}
\end{align}
Let $\Phi_{\rm K}({\bf r})$ be the solution 
for the case $\phi^{\rm q}(x)=0$. 
Then $\Phi_{\rm K}({\bf r})$ satisfies the following energy eigenequation:
\begin{align}
 v_{\rm F} \left[
 \sigma_x \hat{p}_x + \sigma_y (\hat{p}_y + eBx)
 \right] \Phi_{\rm K}({\bf r}) = E \Phi_{\rm K}({\bf r}).
\end{align}
The solutions are the Landau states specified by integer $n$ and a center
coordinate $X$ as [see Eq.~(\ref{eq:LL})]
\begin{align}
 \Phi^{\rm LL}_{nX}({\bf r}) 
 &= C_{nX} e^{-i\frac{Xy}{{\it l}^2}}
 e^{-\frac{1}{2}\left(\frac{x-X}{\it l}\right)^2} \nn \\
 & \times 
 \begin{pmatrix}
  {\rm sgn}(n)\sqrt{2|n|}H_{|n|-1}\left((x-X)/{\it l}\right) \cr
  -iH_{|n|}\left((x-X)/{\it l}\right)
 \end{pmatrix}.
\end{align}
Applying the parity transformation ${\bf r}\to -{\bf r}$ to 
$\Phi^{\rm LL}_{{\rm K},nX}({\bf r})$, we obtain
\begin{align}
 \Phi^{\rm LL}_{{\rm K},nX}({\bf -r}) = 
 (-1)^{n+1} \sigma_z \Phi^{\rm LL}_{{\rm K},n-X}({\bf r}).
 \label{eq:llp}
\end{align}
The matrix $\sigma_z$ on the right-hand side can be understood by
applying the parity transformation ${\bf r}\to -{\bf r}$ to 
this energy eigenequation:
\begin{align}
 v_{\rm F} \left[
 \sigma_x \hat{p}_x + \sigma_y (\hat{p}_y + eBx)
 \right] \Phi_{\rm K}(-{\bf r}) = -E \Phi_{\rm K}(-{\bf r}).
\end{align}
The negative sign in front of the right-hand side shows that
the energy eigenvalue of $\Phi_{\rm K}(-{\bf r})$ 
is opposite to that of $\Phi_{\rm K}({\bf r})$.
By substituting Eq.~(\ref{eq:llp}) into Eq.~(\ref{eq:armcy}), 
we obtain 
\begin{align}
 \Phi^{\rm LL}_{nX}({\bf r}) =
 \begin{pmatrix}
  \Phi^{\rm LL}_{{\rm K},nX}({\bf r}) \cr -ie^{ig} \sigma_y \Phi^{\rm LL}_{{\rm K},n-X}({\bf r})
 \end{pmatrix}.
\end{align}
By repeating the same argument given in the previous subsection, 
the following standing wave solutions are obtained:
\begin{align}
 \Psi^{\rm LL}_{nX\pm}(x) = N(x) 
 \begin{pmatrix}
  \Phi^{\rm LL}_{{\rm K},nX}({\bf r})
  \cr
  \pm \sigma_y \Phi^{\rm LL}_{{\rm K},n-X}({\bf r})
 \end{pmatrix}.
\end{align}
There are no constraints for the value of $X$.
It is then a straightforward calculation to check that 
$\tau_1(x)$ vanishes at the armchair edge.

\section{Discussion and Summary}\label{sec:dis}

A realistic graphene edge may be a mixture of zigzag and armchair edges.~\cite{klusek00,giunta01,kobayashi05,niimi05} 
The construction of the standing wave near the general edge is 
one of the interesting applications for our framework. 
We believe that 
the Hamiltonian in Eq.~(\ref{eq:H}) can describe the low-energy
electrons in a graphene plane with a general edge.
However, note that this issue is related to the coherence length of the
standing wave. 
In the present paper, we have not considered perturbations 
that break coherence, such as electron-phonon interaction.
Interestingly, the electron-phonon interaction can also be represented as
a deformation-induced gauge field.~\cite{sasaki08ptps,sasaki10-physicaE}
Thus, the gauge field description for the graphene edge may be useful 
when we consider such issues. 

The effective-mass model of Eq.~(\ref{eq:H}) is equivalent to
a chiral gauge theory for graphene that has been proposed by Jackiw and
Pi.~\cite{jackiw07} 
Indeed, by applying $\sigma_x$ to $\Psi_{\rm K'}({\bf r})$ 
in Eq.(\ref{eq:hwave}), 
the Hamiltonian in Eq.~(\ref{eq:H}) may be rewritten as 
\begin{align}
 H' = v_{\rm F} 
 \begin{pmatrix}
  \bsigma \cdot ({\bf {\hat p}}+{\bf A}^{\rm q}({\bf r})) 
  & \phi^{\rm q}({\bf r}) \cr
  \phi^{\rm q}({\bf r})^* & 
  -\bsigma \cdot ({\bf {\hat p}}-{\bf A}^{\rm q}({\bf r})) 
 \end{pmatrix},
 \label{eq:DEq}
\end{align}
which is the electronic Hamiltonian of the chiral gauge theory.
They have investigated zero-mode solutions of the Hamiltonian
with a topological vortex for ${\bf A}^{\rm q}({\bf r})$ 
on the background of Kekul\'e distortion for $\phi^{\rm q}({\bf r})$,
in the context of fractionalization of quantum number.~\cite{hou07,chamon08,chamon08-prb}
Our trial is then to study the graphene edge as a chiral gauge theory,
although our results in this paper do not clarify fully the topological features
of the graphene edge.
It is interesting to note that 
one may find an advantage of a chiral gauge theory 
when we consider the real spins of the electrons.
For example, the magnetism of the edge states
may be understood as a parity anomaly phenomenon.~\cite{semenoff84,sasaki08jpsj}
The various field-theoretical techniques may be utilized
to explore the electronic properties near the edge.
Note also that the perturbation which mixes the electrons in the two
valleys has been examined in the studies on the topological defect in
graphene.~\cite{gonzalez92,lammert00}

We have taken into account the edge as a part of the Hamiltonian.
This strategy stems from the tight-binding lattice model, in which 
the edge is automatically included as a part of the Hamiltonian.
A similar concept is found in the article by 
Berry {\it et al.}~\cite{berry87}, 
in which the authors modeled the edge
using a mass term, $V({\bf r}) \sigma_z$.
They considered that 
a singularity of the mass $V({\bf r}) \to \infty$ outside of the edge 
is necessary, in order to uniquely specify the pseudospin.
We have observed a similar situation for the deformation-induced gauge field 
for the edge, that is, the field is singular at the edge.
It is also interesting to note that $\phi^{\rm q}({\bf r})$ in
Eq.~(\ref{eq:DEq}) corresponds to the mass of a Dirac fermion, 
and that the armchair edge is a singular point as for the mass.

In summary, we have proposed a framework in which 
the edge is represented as 
the deformation-induced gauge field. 
We have used the framework to investigate the standing waves and edge
states in the presence of a mass term and a magnetic field.
The description of the edge using the deformation-induced gauge field
is one attempt to better understand the edge.
If we can describe the variety of edge structures as 
different configurations of a single gauge field, 
it provides a basis to further explore the properties near the edge.

\section*{Acknowledgment}

This work was financially supported by 
a Grant-in-Aid for Specially Promoted Research
(No.~20001006) from the Ministry of Education, Culture, Sports, Science and Technology (MEXT).

\appendix

\section{Rotation of Pseudospin}\label{app:rot-pspin}

The configurations of the pseudospin field for 
three equivalent corners of the graphene BZ
are not the same, as shown in Fig.~\ref{fig:pspin}.
Consideration of this pseudospin behavior is given in this Appendix.

The tight-binding Hamiltonian 
can be written as~\cite{sasaki08ptps}
\begin{align}
 H({\bf k})= -\gamma_0 
 \begin{pmatrix}
  0 & \sum_a f_a({\bf k}) \cr
  \sum_a f^*_a({\bf k}) & 0 
 \end{pmatrix},
 \label{eq:Hk}
\end{align}
where 
$f_a({\bf k}) \equiv e^{i{\bf k}\cdot {\bf R}_a}$ ($a=1,2,3$).
Note that $f_a({\bf k})$ satisfies 
\begin{align}
 f_a({\bf k}+n{\bf b}_1+m{\bf b}_2) = 
 f_a({\bf k}) e^{-i\frac{2\pi}{3} \left(n+m\right)},
 \label{app:fa}
\end{align}
where $n$ and $m$ are integers.
Hence, the representations of $H({\bf k})$, 
$H({\bf k}+{\bf b}_1)$, and $H({\bf k}+{\bf b}_1+{\bf b}_2)$ 
are different from each other, and are related via
$H({\bf k}+{\bf b}_1)=MH({\bf k})M^{-1}$ and 
$H({\bf k}+{\bf b}_1+{\bf b}_2)=M^{-1}H({\bf k})M$, where
\begin{align}
 M = 
 \begin{pmatrix}
  e^{+i2\pi/3} & 0 \cr
  0 & e^{-i2\pi/3}
 \end{pmatrix}
 = \exp\left(i\frac{2\pi}{3}\sigma_z\right).
\end{align}
For the solution $\Psi$ of $H({\bf k})$, 
we have the corresponding solution of the effective Hamiltonian
at ${\bf k}+{\bf b}_1$ as $\Psi_{{\bf b}_1}=M \Psi$.
$M$ is a rotational matrix for the pseudospin
around the $z$-axis, so that the pseudospin of $\Psi$ and 
that of $M \Psi$ are related by rotation around the $z$-axis
by an angle of $2\pi/3$.
This explains why the configurations of the pseudospin field 
around the three equivalent K (K$'$) points are different from each other,
as shown in Fig.~\ref{fig:pspin}.

Next, we consider the effective Hamiltonians for three equivalent K (K$'$) points.
By expanding $f_a({\bf k})$ around the wave vector of the K point, 
${\bf k}_{\rm F}= (4\pi/3a,0)$, we obtain
$f_a({\bf k}_{\rm F}+{\bf k}) = f_a({\bf k}_{\rm F}) 
 + if_a({\bf k}_{\rm F}) {\bf k}\cdot {\bf R}_a + \cdots$.
Using $f_1({\bf k}_{\rm F})=1$,
$f_2({\bf k}_{\rm F})=e^{-i\frac{2\pi}{3}}$, and
$f_3({\bf k}_{\rm F})=e^{+i\frac{2\pi}{3}}$,
we have $H({\bf k}_{\rm F}+{\bf k}) = v_{\rm F} \bsigma \cdot {\bf
p}+\cdots$, where ${\bf p}=\hbar {\bf k}$ and 
$v_{\rm F}=3\gamma_0 a_{\rm cc}/2\hbar$.
Then
$H({\bf k}_{\rm F}+{\bf b}_1+{\bf k}) = M v_{\rm F} \bsigma \cdot {\bf p} M^{-1}+\cdots$,
and $H({\bf k}_{\rm F}+{\bf b}_1+{\bf b}_2+{\bf k}) = 
M^{-1} v_{\rm F} \bsigma \cdot {\bf p} M+\cdots$ are obtained.
The same argument can be applied to the K$'$ points. 
For the K$'$ point at $-{\bf k}_{\rm F}$, 
we obtain the effective Hamiltonian 
$H(-{\bf k}_{\rm F}+{\bf k}) = v_{\rm F} \bsigma' \cdot {\bf p}+\cdots$.
It is then straightforward to obtain
$H(-{\bf k}_{\rm F}+{\bf b}_1+{\bf k}) = M v_{\rm F} \bsigma' \cdot {\bf
p} M^{-1} +\cdots$, and
$H(-{\bf k}_{\rm F}+{\bf b}_1+{\bf b}_2+{\bf k}) = M^{-1} v_{\rm F} \bsigma' \cdot {\bf p} M+\cdots$.
This difference in the representations of the effective Hamiltonians
does not cause a problem, 
because a coordinate transformation 
can be used to eliminate the $M$ matrix from one effective Hamiltonian
(see also Appendices in Ref.~\onlinecite{sasaki08_chiral}).~\cite{slonczewski58}
Here, we imply the coordinate transformation 
as the rotation of the $x$ and $y$-axes by $\pm 2\pi/3$.
A coordinate transformation cannot alter the physics, and therefore
the physical result derived from the effective Hamiltonians are the
same. 
Rather, by using the change of the effective Hamiltonians 
under a translation given by the reciprocal lattice vectors, 
a constraint for the form of the effective Hamiltonians can be obtained.
For example, 
the deformation Hamiltonian, $\bsigma \cdot {\bf A}^{\rm q}({\bf r})$,
should transform in the same way as $\bsigma \cdot {\bf p}$.
Therefore, we must have 
$H({\bf k}_{\rm F}+{\bf k}) = v_{\rm F} \bsigma \cdot ({\bf p}+{\bf
A}^{\rm q}) +\cdots$, 
$H({\bf k}_{\rm F}+{\bf b}_1+{\bf k}) = M v_{\rm F} \bsigma \cdot ({\bf
p}+{\bf A}^{\rm q}) M^{-1} +\cdots$,
and 
$H({\bf k}_{\rm F}+{\bf b}_1+{\bf b}_2+{\bf k}) = 
M^{-1} v_{\rm F} \bsigma \cdot ({\bf p}+{\bf A}^{\rm q}) M+\cdots$.
Otherwise, there would be three physically distinct effective Hamiltonians
for the same K point. 
The deformation-induced gauge field ${\bf A}^{\rm q}({\bf r})$
satisfies this constraint, because
\begin{align}
 v_{\rm F}(A_x^{\rm q}({\bf r}) -i A_y^{\rm q}({\bf r})) = 
 \sum_a \delta \gamma_{0,a}({\bf r}) f_a({\bf k}_{\rm F}).
\end{align}
Note that this equation is equivalent to Eq.~(\ref{eq:gauge}).
The phase factor of $e^{\mp i 2\pi/3}$ appears
when we change ${\bf k}_{\rm F}$ to ${\bf k}_{\rm F}+{\bf b}_1$ 
and to ${\bf k}_{\rm F}+{\bf b}_1+{\bf b}_2$, due to the factor of $f_a({\bf k}_{\rm F})$ on the right-hand side.
A notable feature is that 
the constraint must be satisfied 
for a strong lattice deformation 
that corresponds to a large value of ${\bf A}^{\rm q}({\bf r})$. 
Therefore, the direction of the gauge field does not change,
although the values of $\gamma_{0,a}({\bf r})$ are renormalized 
for a strong deformation.

\section{Edge states and Mass}\label{app:edge}

The edge states in the presence of a mass term 
is of interesting,
because the magnetism of the edge states is related to the
generation of a local spin-dependent mass term 
due to the coulombic interaction.~\cite{sasaki08jpsj}
Here, we show how to obtain the edge states
in the presence of a uniform mass term.

By substituting Eq.~(\ref{eq:phiR}) into 
$H^m_{\rm K}(y) \Psi_{\rm K}(y) = E \Psi_{\rm K}(y)$, 
we obtain instead of Eq.~(\ref{eq:weyl+g})
\begin{align}
\begin{split}
 & p_x + A^{\rm q}_x(y) + \hbar \frac{dg(y)}{dy} = D \cosh(2g(y)+f), \\
 & \hbar \frac{d}{dy}\left(\frac{|y|}{\xi}\right) = D \sinh(2g(y)+f), 
\end{split}
\label{eq:G}
\end{align}
where the variables $D$ and $f$ 
are respectively defined as
\begin{align}
 D \equiv \pm \frac{1}{v_F} \sqrt{E^2-m^2} \ {\rm and} \ \tanh(f) \equiv - \frac{m}{E}.
 \label{eq:X}
\end{align}
The solution of the second equation in~(\ref{eq:G}) is 
\begin{align}
 2g(y) + f= 
 \begin{cases}
  \displaystyle - \sinh^{-1} \left( \frac{\hbar}{\xi D} \right) & (y < 0), \\
  \displaystyle + \sinh^{-1} \left( \frac{\hbar}{\xi D} \right) & (y > 0).
 \end{cases}
 \label{eq:caseg}
\end{align}
The first equation in Eq.~(\ref{eq:G}) is integrated
with respect to $y$ from $-\xi_g$ to $\xi_g$.
Considering the limit $\xi_g \to 0$, 
only singular functions of $A_x^{\rm q}(y)$
and $g(y)$ at $y=0$ can survive
after the integration, so that we obtain
\begin{align}
 - \sinh^{-1} \left( \frac{\hbar}{\xi D} \right)
 = \int_{-\xi_g}^{\xi_g} A^{\rm q}_x(y)dy.
 \label{eq:gauge_con}
\end{align}
Using Eqs.~(\ref{eq:caseg}) and (\ref{eq:gauge_con}),
we see from the first equation in ~(\ref{eq:G}) that 
\begin{align}
 D =
 \frac{p_x}{\cosh\left(\int_{-\xi_g}^{\xi_g} A^{\rm q}_x(y)dy \right)} 
 \label{eq:Xs}
\end{align}
holds except very close to the edge.
From Eqs.~(\ref{eq:gauge_con}) and (\ref{eq:Xs}),
we see that $\xi$ is given by 
\begin{align}
 \frac{1}{\xi}
 = -k_x \tanh\left(\int_{-\xi_g}^{\xi_g} A^{\rm q}_x(y)dy \right).
\end{align}
Note that $\xi$ in the presence of a mass term
is identical to $\xi$ in the absence of the mass given in 
Eq.~(\ref{eq:xi-n}).
The mass term would affect $\xi$, but this is not the case.
From Eqs.~(\ref{eq:X}) and (\ref{eq:Xs})
we obtain the energy eigenvalue 
\begin{align}
 E=\pm \sqrt{m^2+(v_FD)^2}.
 \label{eq:Ene}
\end{align}
When $\int_{-\xi_g}^{\xi_g} A^{\rm q}_x(y)dy \to \infty$, we obtain 
$D = 0$ and $E=\pm|m|$.

According to the definition of $f$ in Eq.~(\ref{eq:X}), 
the sign of $f$ depends on the signs of both $m$ and $E$.
Let us first consider the case of $E<0$, by which
we have $f={\rm sign}(m)|f|$.
Using this expression for $f$ in Eq.~(\ref{eq:caseg}),
we obtain
\begin{align}
 g(y) =
 \begin{cases}
  \displaystyle + \frac{1}{2} \int_{-\xi_g}^{\xi_g} 
  A^{\rm q}_x(y)dy - \frac{1}{2}{\rm sign}(m) |f| &(y < 0), \\
  \displaystyle - \frac{1}{2} \int_{-\xi_g}^{\xi_g} 
  A^{\rm q}_x(y)dy - \frac{1}{2}{\rm sign}(m) |f| &(y > 0).
 \end{cases}
 \label{eq:caseg(y)}
\end{align}
To determine $|f|$,
we substitute Eq.~(\ref{eq:Ene}) into Eq.~(\ref{eq:X}),
and considering that $\tanh(|f|)$ can be approximated as 
$1-2e^{-2|f|}$ for $|f|\gg 1$,
we then have 
$|f| \approx \left| \int_{-\xi_g}^{\xi_g} A^{\rm q}_x(y)dy \right|$
for $\left| \int_{-\xi_g}^{\xi_g} A^{\rm q}_x(y)dy \right|\gg 0$.
Since $\int_{-\xi_g}^{\xi_g} A^{\rm q}_x(y)dy \gg 0$ 
for the zigzag edge, 
Eq.~(\ref{eq:caseg(y)}) becomes
\begin{align}
 g(y) \approx 
 \begin{cases}
  \displaystyle \int_{-\xi_g}^{\xi_g} A^{\rm q}_x(y)dy &(y < 0), \\
  \displaystyle 0 \ &(y > 0),
 \end{cases}
 \label{eq:gup}
\end{align}
when $m < 0$.
The wave function of this eigenstate for $y>0$,
which has unpolarized pseudospin, 
is negligible due to the normalization.
Thus, the localized state with energy $E=-|m|$ 
in the valence energy band
can appear near the edge only for $y<0$, 
and the wave function is given by 
$\Psi_{\rm K}(y<0) \propto \exp(-|y|/\xi){}^t(1,0)$.
Similarly, for the case of $m > 0$, we have
\begin{align}
 g(y) \approx 
 \begin{cases}
  \displaystyle 0 &(y < 0), \\
  \displaystyle -\int_{-\xi_g}^{\xi_g} A^{\rm q}_x(y')dy' &(y > 0).
 \end{cases}
 \label{eq:gdn}
\end{align}
The corresponding wavefunction is has the pseudospin down state, 
which appears only for $y>0$ near the edge.
It is noted that the mass term automatically selects the region 
where the edge state can appear, $y>0$ or $y<0$. 
This is reasonable, because we have used the particle-hole symmetry
operator $\sigma_z$ to restrict the edge state 
only for $y>0$ or $y<0$ in Sec.~\ref{ssec:es}. 
The particle-hole symmetry operator is nothing but the mass term.

\bibliographystyle{apsrev}

\end{document}